\documentclass[10pt]{JHEP3} 

\usepackage{epsfig,multicol,bbm}

\newcommand\fverb{\setbox\fverbbox=\hbox\bgroup\verb}
\newcommand\fverbdo{\egroup\medskip\noindent%
            \fbox{\unhbox\fverbbox}\ }
\newcommand\fverbit{\egroup\item[\fbox{\unhbox\fverbbox}]}
\newbox\fverbbox

\usepackage{amsmath}
\usepackage{amssymb}
\usepackage{amsfonts}
\usepackage{latexsym}
\usepackage{graphicx}
\usepackage{dcolumn}
\usepackage{bm}
\usepackage{empheq}
\usepackage{times}  
\newcommand{\eq}[1]{\begin{equation}#1\end{equation}}

\newcommand{\ea}[1]{\begin{equation}\begin{aligned}#1\end{aligned}\end{equation}}
\newcommand{\itm}[1]{\begin{itemize}#1\end{itemize}}
\newcommand{\od}[2]{\frac{\textrm{d} #1}{\textrm{d} #2}}  
\newcommand{\pd}[2]{\frac{\partial #1}{\partial #2}}  
\newcommand{\lrp}[1]{\left( #1 \right)}  
\newcommand{\lrsb}[1]{\left[ #1 \right]}  
\newcommand{\lrcb}[1]{\left\{ #1 \right\}}  
\newcommand{\lrab}[1]{\left\langle #1 \right\rangle}  
\newcommand{\ab}[1]{\langle #1 \rangle} 
\newcommand{\lrmb}[1]{\left| #1 \right| } 
\newcommand{\soev}[3]{\lrab{ #1 \lrmb{ #2}#3 }} 
\def\la{\langle}
\def\ra{\rangle}

\def\tr{\mathrm{tr}}
\def\rd{\partial}
\def\vx{\bm{x}}
\def\vk{\bm{k}}

\def\mpl{M_{\textrm{pl}}}

\def\dep{\delta\phi}
\def\dop{\dot{\phi}}

\def\qsg{Q_{\sigma}}
\def\qs{Q_{s}}
\def\I{\textrm{I}}
\def\doq{\dot{Q}}
\def\ca{c_{\textrm{a}}}
\def\ce{c_{\textrm{e}}}
\def\Rc{\mathcal{R}}
\def\Sc{\mathcal{S}}

\title{Primordial Non-Gaussianities from the Trispectra in Multiple Field Inflationary Models}

\author{Xian Gao\\
    Key Laboratory of Frontiers in Theoretical Physics,\\
    Kavli Institute for Theoretical Physics, Chinese Academy of
    Sciences, Beijing 100190,
    P.R.China\\
    E-mail: \email{gaoxian@itp.ac.cn}
    }

\author{Miao Li\\
    Key Laboratory of Frontiers in Theoretical Physics,\\
    Kavli Institute for Theoretical Physics, Chinese Academy of
    Sciences, Beijing 100190,
    P.R.China\\
    E-mail: \email{mli@itp.ac.cn}
    }

\author{Chunshan Lin\\
    The Interdisciplinary Center for Theoretical Study,\\
     University of Science and Technology of China, Hefei, Anhui 230026, P.R.China\\
     Key Laboratory of Frontiers in Theoretical Physics,\\
    Kavli Institute for Theoretical Physics, Chinese Academy of
    Sciences, Beijing 100190,
    P.R.China\\
    E-mail: \email{lics@mail.ustc.edu.cn}
    }

\preprint{CAS-KITPC/ITP-119\\
            USTC/ICTS-09-06}

\abstract{
    We investigate the primordial non-Gaussianities from the trispectra in multi-field inflation models,
    which can be seen as generalization of multi-field $k$-inflation and multi-DBI inflation. We derive
    the full fourth-order perturbation action for the inflaton fields and evaluate the four-point correlation
    functions for the perturbations in the limit $\ca
\ll 1$ and $\ce \ll1$. There are three types of momentum-dependent
shape functions which arise from three types of four-point
interaction vertices.
    The final trispectrum of the curvature perturbation can be expressed in terms of
    the deformations and permutations of these three shape functions,
    and is determined by $\ca$, $\ce$, $\lambda$ , $\Pi$ which depend
    on the non-linear structure of the model and also the transfer
    function $T_{\Rc\Sc}$.
    We also discuss the parameter space for the trispectrum and plot
    the shape diagrams for the trispectrum both for visualization and for distinguishing different
    shapes from each other.
    }


\begin{document}

\section{Introduction}

Current observational data support cosmological inflation greatly
\cite{Komatsu:2008hk}, in which primordial perturbations assumed
responsible for Cosmic Microwave Background anisotropies and
Large-Scale Structure formation are generated from quantum
fluctuations and stretched to superhorizon scales during inflation
(see e.g. \cite{Lyth:1998xn} for a review). Actually, inflation
itself is not a single model, but rather a theoretical framework.
One of the most robust predictions of inflation is the almost
scale-invariant, Gaussian and adiabatic primordial fluctuation.
However, from the point of view of power spectrum (i.e., Fourier
transformation of the tree-level two-point equal-time correlation
function) of the cosmological perturbation, many inflation models
are ``degenerate". The theory and observation of power spectrum do
not give us an unique theory of inflation. Phenomena beyond
linear-order have been extensively investigated over the past
several years.

The most significant progress beyond power spectrum is the
investigation of statistical non-Gaussianities of CMB anisotropies
and primordial fluctuations \cite{Komatsu:2009kd} (see e.g.
\cite{Bartolo:2004if} for a nice review). From the field theoretical
point of view, non-Gaussianity describes \emph{interactions} among
perturbations, which will cause non-vanishing higher-order
correlation functions. Such interactions are mandatory in any
realistic inflationary models, which come from both the non-linear
nature of gravitation and the self-interactions of inflation
field(s). In standard slow-roll inflation scenario, however, the
non-Gaussianities have been proved too small to be detected
\cite{Maldacena:2002vr} (see also \cite{Acquaviva:2002ud}), even
with PLANCK satellite \cite{PLANCK}. Thus, any detection of
non-Gaussianities would not only rule out the simplest inflation
models but also give us valuable insight into fundamental physics of
inflation
\cite{Boubekeur:2005fj,Yadav:2007yy,Smith:2009jr,Senatore:2009gt,Curto:2009pv,Nitta:2009jp,Rossmanith:2009cy,Sefusatti:2009xu,Gong:2009dt}.

Generally speaking, primordial non-Gaussianities can be large, if
one or more of the following four conditions are violated: 1)
slow-roll conditions, 2) canonical kinetic terms, 3) single field
and 4) Bunch-Davies vacuum \cite{Bartolo:2004if}. For single-field
models, for example, various possibilities have been investigated in
order to generate large non-Gaussianities by introducing complicated
kinetic terms
\cite{Seery:2005wm,Chen:2006nt,Huang:2006eh,Cheung:2007st,Chen:2006xjb,Chen:2008wn,Seery:2006vu,Seery:2006js,Byrnes:2006vq,Arroja:2008ga,Seery:2008ax,Chen:2009bc,Arroja:2009pd}
which belong to the more general class of $k$-inflation models
\cite{ArmendarizPicon:1999rj,Garriga:1999vw}, in which the
small-speed of sound $c_s\ll 1$ enhances the derivative-coupling of
perturbations, or other mechanisms which enhance the interactions
during inflation or non-linearities during superhorizon evolution
\cite{Bernardeau:2002jf,Dvali:2003em,Creminelli:2003iq,Lyth:2005qk,Alabidi:2006wa,Suyama:2008nt,Cogollo:2008bi,Rodriguez:2008hy,Byrnes:2008zz,Holman:2007na,Meerburg:2009ys}.

From the point of view of data analysis, there are two typical types
of non-Gaussianities which are most interesting: the so-called
``local" type and the ``equilateral" type. The former describes the
strength of modulations of short wavelength perturbation modes by
long wavelength modes, while the later describes the correlations
among
 modes with similar wavelengthes. It has been clear from the studies of non-Gaussianities in
 single-field models that, single-field inflationary models with non-canonical kinetic terms can have large ``equilateral"-type
 non-Gaussianities, since the ``derivative-coupled" interactions among different modes are enhanced by the non-canonical structure
 of the kinetic term. While single-field models can
 never generate large ``local"-type non-Gaussianities, unless the generalized slow-roll conditions are abandoned.

Multiple field inflationary models provide us with more
possibilities to generate large non-Gaussianities during inflation
\cite{Langlois:2008qf,Langlois:2008wt,Langlois:2009ej,Gao:2008dt,Gao:2009gd,Arroja:2008yy,Kawakami:2009iu,Mizuno:2009cv,Lehners:2009ja,Bartolo:2001cw,Seery:2005gb,Bernardeau:2002jy,Wands:2002bn,Byrnes:2008wi,Byrnes:2008zy,Langlois:2008vk,Kawasaki:2008sn,Langlois:2008mn,Rigopoulos:2005xx,Ji:2009yw,Pi:2009an}.
It has been shown that multi-inflaton models with non-canonical
kinetic terms are also mainly characterized by equilateral-type
non-Gaussianities rather than local-type. Local-type
non-Gaussianities can be generated during superhorizon evolution of
cosmological perturbations, for example in inhomogeneous
``end-of-inflation" models
\cite{Alabidi:2006hg,Sasaki:2008uc,Naruko:2008sq,Huang:2009vk,Huang:2009xa}
or curvaton scenario
\cite{Sasaki:2006kq,Malik:2006pm,Huang:2008bg,Huang:2008rj,Huang:2008ze,Li:2008fma,Ichikawa:2008iq,Kobayashi:2009cm}.

In this work, we investigate primordial non-Gaussianities from the
trispectra (i.e. Fourier transformation of equal-time four-point
correlation functions) of cosmological perturbations in general
multiple field models. We consider the model with general
scalar-fields Lagrangian of the form $P(X^{IJ},\phi^I)$. This form
of Lagrangian was first investigated in
\cite{Langlois:2008wt,Arroja:2008yy}. It includes single-field
models with non-canonical kinetic term
\cite{ArmendarizPicon:1999rj,Garriga:1999vw}, multi-field
k-inflation \cite{Langlois:2008mn,Gao:2008dt} and multi-field DBI
models
\cite{Langlois:2008qf,Langlois:2008wt,Gao:2009gd,Mizuno:2009cv} as
special cases, and thus deserves detailed studies. Bispectrum in
general multi-field models with small speed of sound has been
investigated
 in
\cite{Langlois:2008qf,Langlois:2008wt,Gao:2008dt,Arroja:2008yy}.
Trispectra in single-field models have been investigated by various
author
\cite{Huang:2006eh,Seery:2006vu,Seery:2006js,Byrnes:2006vq,Arroja:2008ga,Seery:2008ax,Chen:2009bc,Arroja:2009pd}
and in multi-field models until very recently
\cite{Gao:2009gd,Mizuno:2009cv,Lehners:2009ja,Byrnes:2009qy}.

We start from the the full fourth-order action for the scalar
perturbations and then evaluate the leading-order four-point
functions around sound-horizon crossing in the approximation $\ca
\ll 1$ and $\ce \ll 1$, where $\ca$ and $\ce$ are propagation speeds
of adiabatic and entropy modes respectively. In this work we
restrict the calculations to the contributions from the so-called
``contact interaction", i.e. the contributions from the ``four-point
direct interaction vertices"{\footnote{Obviously the tree-level
four-point correlation functions have two origins: one is the
four-point vertices, the other is correlating two three-point
vertices, e.g. with scalar modes
\cite{Gao:2009gd,Chen:2009bc,Mizuno:2009cv} or graviton
\cite{Seery:2008ax}. A full analysis should include these two
contributions together.}}. As in the case in general single-field
models, in the leading-order, we find three fundamental shape
functions $A_1(k_1,k_2,k_3,k_4)$, $A_2(k_1,k_2;\vk_1,\vk_2)$ and
$A_3(\vk_1,\vk_2;\vk_3,\vk_4)$, corresponding to three different
types of four-point interaction vertices. Moreover, in addition to
the pure adiabatic four-point function $\lrab{\qsg\qsg\qsg\qsg}$,
there also exist pure entropy four-point function{\footnote{This is
different from the case of bispectrum. There is no leading-order
pure entropy three-point correlation $\lrab{\qs\qs\qs}$ and the
mixed contribution is of the form $\lrab{\qsg\qs\qs}$, see
\cite{Langlois:2008qf,Langlois:2008wt,Gao:2008dt,Arroja:2008yy} for
details.}} $\lrab{\qs\qs\qs\qs}$ and one mixed contribution
$\lrab{\qsg\qsg\qs\qs}$ (we denote $\qsg$ and $\qs$ as the adiabatic
and entropy modes respectively). The four-point correlation
functions $\lrab{\qsg\qsg\qsg\qsg}$, $\lrab{\qsg\qsg\qs\qs}$ and
$\lrab{\qs\qs\qs\qs}$ can be generally written in terms of various
deformations and permutations of these three types of shape
functions. Since the observable is the curvature perturbation $\Rc$,
we also investigate the trispectrum of the curvature perturbation
$\lrab{\Rc\Rc\Rc\Rc}$ on large scales.

In this work, we assume the propagation speeds of adiabatic mode and
entropy modes be different: $\ca \neq \ce$. Thus, unlike the
multi-field DBI models
\cite{Langlois:2008qf,Langlois:2008wt,Gao:2009gd,Mizuno:2009cv}
where $\ca = \ce =c_s$ and we can abstract common shape functions
for both $\lrab{\qsg\qsg\qsg}$ and $\lrab{\qsg\qs\qs}$, in this work
we will see that  $\ca$, $\ce$ enter the definition of the shape
function intrinsically. More precisely, as we will see, the
trispectrum of the curvature perturbation is determined by four
parameters $\ca$, $\ce$, $\lambda$, $\Pi$ which arise from the
non-linear structure of the theory and also $T_{\Rc\Sc}$ which
characterizes the transfer from entropy modes to adiabatic modes
during superhorizon evolution.

This paper is organized as follows. In section 2, we describe the
model and set up the perturbation theory in spatially-flat gauge,
then we briefly review the linear perturbation and the power spectra
in section 3. In section 4, we give the full fourth-order
perturbation action and derive the interaction Hamiltonian in the
interaction picture. And we calculate the four-point functions and
the trispectra in section 5. In section 6, we give a general
discussion on characterizing the trispectrum and we plot the shape
diagrams and investigate the non-linear parameters. In the end we
make a conclusion and discuss some related issues.

In this paper, we choose $c=\hbar=\mpl\equiv \frac{1}{8\pi G} = 1$.

\section{Basic Setup}

\subsection{Model and Background}

In this work we consider a general class of multi-field models
containing $\mathcal{N}$ scalar fields coupled to Einstein gravity.
The action takes the form
    \eq{{\label{action_original}}
        S = \int d^4x\, \sqrt{-g} \lrsb{ \frac{R}{2} +
        P \lrp{X^{IJ},\phi^I}} \,,
        }
where $\phi^I$ ($I=1,2,\cdots,\mathcal{N}$) are scalar fields
acting as inflaton fields, and
    \eq{
        X^{IJ} \equiv -\frac{1}{2} g^{\mu\nu} \rd_{\mu} \phi^I \rd_{\nu}
        \phi^J \,,
        }
is the kinetic term (matrix), $g_{\mu\nu}$ is the spacetime metric
tensor with signature $(-,+,+,+)$. ``$I,J$"-indices are raised,
lowered and contracted by $\mathcal{N}$-dimensional field-space
metric $G_{IJ} = G_{IJ}(\phi^I)$. This form of Lagrangian
includes multi-field k-inflation and multi-DBI models as special
cases{\footnote{For example, multi-field k-inflation has the
scalar-field Lagrangian as $P(X,\phi^I)$, where $X \equiv \tr X^{IJ}
= G_{IJ} X^{IJ}$. While in multi-field DBI models, $P(X^{IJ},\phi^I)
= - \frac{1}{f(\phi^I)} \lrp{ \sqrt{ \mathcal{D} }-1} - V(\phi^I)$
with $ \mathcal{D} = 1 -2f G_{IJ}X^{IJ} + 4f^2 X^{[I}_I X^{J]}_J -
8f^3 X^{[I}_I X^J_J X^{K]}_K + 16 f^4 X^{[I}_I X^J_J X^K_K
X^{L]}_L$. See later discussion for details. }}.

Modern ``action approach" of cosmological perturbations is based on
the ADM formalism of gravitation, in which spacetime metric is
written as
    \eq{
        ds^2 = -N^2 dt^2 +h_{ij}(dx^i+N^idt)(dx^j+N^jdt)\;,
    }
with $N=N(t,\vx)$ is the lapse function and $N_i=N_i(t,\vx)$ is the
shift vector, $h_{ij}$ is the spatial metric on constant time
hypersurfaces. The ADM formalism is convenient because the equations
of motion for $N$ and $N^i$ are exactly the energy and momentum
constraints which are easy to solve. Under the ADM formalism, the
action (\ref{action_original}) can be written as (up to total
derivative terms)
    \eq{{\label{action_einstein}}
    S = \int dtd^3x\, \sqrt{h} N \lrp{ \frac{1}{2}R^{(3)} + \frac{1}{2N^2} \lrp{ E_{ij}E^{ij}-E^2 } } +
    \int dtd^3x\, \sqrt{h}N\,P  \,,
    }
where $h\equiv \det h_{ij}$ and the symmetric tensor \eq{
    E_{ij} \equiv \frac{1}{2} \lrp{ \dot{h}_{ij} - \nabla_iN_j - \nabla_jN_i
    } \,,
} with $\nabla_i$ is the spatial covariant derivative defined with
spatial metric $h_{ij}$ and $E \equiv \mathrm{tr}E_{ij} =
h^{ij}E_{ij}$. $R^{(3)}$ is the three-dimensional Ricci scalar which
is computed from the spatial metric $h_{ij}$. In ADM formalism,
spatial indices are raised and lowered using $h_{ij}$ and $h^{ij}$.

In ADM formalism, the kinetic matrix $X^{IJ}$ can be written as
    \eq{
        X^{IJ} = -\frac{1}{2} h^{ij} \rd_i \phi^I \rd_j \phi^J +
        \frac{1}{2N^2} v^Iv^J \,,
    }
where
    \eq{
        v^I \equiv \dop^I - N^i \nabla_i \phi^I \,.
    }

\subsubsection{Equations of Motion}

The equations of motion for the scalar fields are
    \eq{
        \nabla_{\mu} \lrp{ P_{,\ab{IJ} } \rd^{\mu} \phi^I } + P_{,J} = 0\,,
    }
where $\nabla_{\mu}$ is the four-dimensional covariant derivative.
Here and in what follows, we denote
    \eq{
        P_{,\ab{IJ} } \equiv \pd{P}{X^{IJ}}\,,\qquad P_{,\ab{IJ}\ab{KL} }
        \equiv \frac{ \rd^2 P}{ \rd X^{IJ} \rd X^{KL}} \,,
    }
for shorthand.

The equations of motion for $N$ and $N_i$ are Hamiltonian and
momentum constraints respectively,
    \ea{{\label{constraints}}
        R^{(3)} + 2P - \frac{2}{N^2} P_{,\ab{IJ}} v^I v^J -
        \frac{1}{N^2} \lrp{ E_{ij}E^{ij} - E^2 } &=0\,,\\
        \nabla_j \lrp{ \frac{1}{N} \lrp{ E^j_i - E \delta^j_i } } -
        \frac{P_{,\ab{IJ}}}{N} v^I \nabla_i\phi^J &=0 \,.
    }

\subsubsection{Background}

In this work, we investigate scalar perturbations around a flat FRW
background, the background spacetime metric takes the form
    \eq{
        ds^2 = -dt^2 + a^2(t) \delta_{ij} dx^i dx^j \,,
    }
where $a(t)$ is the so-called scale-factor. The Friedmann equation
and the continuity equation are
    \ea{
        H^2 &= \frac{\rho}{3} \equiv \frac{1}{3} \lrp{ 2X^{IJ} P_{,\ab{IJ} } - P
        } \,,\\
        \dot{\rho} &= -3H( \rho + P ) \,.
    }
In the above equations, all quantities are background values. From
the above two equations we can also get another convenient equation
    \eq{{\label{eq_dotH}}
        \dot{H} = -X^{IJ} P_{,\ab{IJ}} \,.
    }
The background equations of motion for the scalar fields are
    \eq{
        P_{,\ab{IJ}} \ddot{\phi}^I + \lrp{ 3HP_{,\ab{IJ}} + \dot{P}_{,\ab{IJ}}
        } \dop^I - P_{,J} = 0\,,
    }
where $P_{,I}$ denotes derivative of $P$ with respect to $\phi^I$:
$P_{,I} \equiv \pd{P}{\phi^I}$.

In this work, we investigate cosmological perturbations during an
exponential inflation period. Thus, from (\ref{eq_dotH}) it is
convenient to define a slow-roll parameter
    \eq{
        \epsilon \equiv -\frac{\dot{H}}{H^2} = \frac{P_{,\ab{IJ}} \dop_0^I
        \dop_0^J}{2H^2} \,.
    }

\subsection{Perturbation Theory in Spatially-flat Gauge}

The scalar metric fluctuations about our background can be written
as (see \cite{MFB,RHBrev1} for nice review of the theory of
cosmological perturbations) 
\ea{
    \delta N  &= \alpha \, , \\
    \delta N_i  &= \rd_i \beta \, , \\
    \delta g_{ij}  &=  - 2 a^2 \bigl( \psi \delta_{ij} - \partial_i \partial_j E) \,
    }
where $\alpha, \beta, \psi$ and $E$ are functions of space and
time{\footnote{This form of ansatz corresponds to $\delta g_{00} =
1- N^2 + N_iN^i$ and $ \delta g_{0i} = N_i $.}}.
 The scalar field perturbations are denoted by $\dep^I \equiv Q^I$.

Before proceeding, we would like to analyze the (scalar) dynamical
degrees of freedom in our system. In the beginning we have
$\mathcal{N}+4$ apparent scalar degrees of freedom. The
diffeomorphism of Einstein gravity eliminates two of
them{\footnote{See \cite{MFB} for a detailed discussion on the gauge
issue of cosmological perturbations.}}, leaving us $\mathcal{N}+2$
scalar degrees of freedom. Furthermore, two of these $\mathcal{N}+2$
degrees of freedom are non-dynamical. In ADM formalism, these are
just the fluctuations $\delta N=\alpha$ and $\delta N_i = \rd_i
\beta$. Thus, there are $\mathcal{N}$ propagating degrees of freedom
in our system. As has been addressed, the diffeomorphism invariance
allows us to choose convenient gauges to eliminate two degrees of
freedom. In single-field models, there are two convenient gauge
choices: comoving gauge corresponding to choosing $\dep=E=0$ or
spatially-flat gauge corresponding to $\psi=E=0$. In multi-field
case, the comoving gauge loses its convenience since we cannot set
$\delta\rho=0$ in multi-field case. Thus, in this work we use the
spatially-flat gauge{\footnote{In the spatially-flat gauge, it is
obvious that the dynamical degrees of freedom come from the scalar
fields perturbations $\dep^I$. In Einstein gravity, the scalar-type
metric perturbations are essentially non-dynamical. While in
non-Einstein gravity, e.g. in recently proposed Ho\v{r}ava gravity,
this may not be the case \cite{Gao:2009ht,Gao:2009bx}}}.

In the spatially-flat gauge, propagating degrees of freedom for
scalar perturbations are the inflaton field perturbations
$Q^I(t,\vx)$, while  $\delta N$ and $\delta N_i$ are non-dynamical
constraints. In this work, we focus on scalar
perturbations{\footnote{In general, it is well-known that in the
higher-order perturbation theories, scalar/vector/tensor
perturbation modes are coupled together. However, from the point of
view of the perturbation action approach, these couplings are
equivalent to exchanging various modes. For example, the effects of
tensor perturbation on scalar modes have been investigated in
\cite{Dimastrogiovanni:2008af} through graviton exchange approach.
In this work, we focus on interactions of scalar modes themselves,
thus we can neglect tensor perturbations.}}. The perturbations take
the form \ea{{\label{pert_ansatz}}
    \phi^I(t,\vx) &= \phi_0^I(t) + Q^I(t,\vx) \,,\\
    h_{ij} &\equiv a^2 \delta_{ij} \,\\
    N &= 1+\alpha_1 + \alpha_2 + \cdots \,,\\
    N_i &= \rd_i (\beta_1 + \beta_2 +\cdots) + \theta_{1i} +
    \theta_{2i} +\cdots \,,
} where $\phi_0^I(t)$ is the background value, and
$\alpha_n,\beta_{n},\theta_{ni}$ are of order $\mathcal{O}(Q^n)$.

The next step is to solve the constraints $\alpha_n$, $\beta_n$ and
$\theta_{ni}$ in terms of $Q^I$. Fortunately, in order to expand the
action to fourth-order in $Q^I$, the solutions for the constraints
up to the second-order are adequate. In general, we need the
solutions for constraints $N$ and $N_i$ up to
$\mathcal{O}(Q^{[n/2]})$ order ($[n/2]$ means the largest integer
$\leq n/2$), if we want to expand the action to $\mathcal{O}(Q^n)$
order and to calculate the $n$-point correlation functions.

\subsubsection{Solving the Constraints}

 At the first-order in $Q^I$, a
particular solution for equations (\ref{constraints}) is:
    \ea{{\label{con_sol_1}}
        \alpha_1 &= \frac{1}{2H} P_{,\ab{IJ}} \dop^I Q^J \,,\\
        \beta_1 &=  \frac{a^2}{2H} \rd^{-2} \left[ \lrp{  P_{,\ab{IJ}} + 2\, X^{KL} P_{,\ab{IJ}\ab{KL}} } \lrp{ \frac{X^{IJ}}{H} P_{,\ab{KL}} \dop^K Q^L  -\dop^I \doq^J} \right. \\
        &\qquad\qquad\qquad\qquad \left. -3H P_{,\ab{IJ}} \dop^I Q^J - P_{,\ab{IJ}K}Q^K 2X^{IJ} +
        P_{,I}Q^I \right] \,,\\
        \theta_{1i} &=0 \,.
    }
Here and in what follows, repeated lower indices are contracted
using $\delta_{ij}$, and $\rd^2 \equiv \rd_i\rd_i$. $\rd^{-2}$ is a
formal notation and should be understood in fourier space.

Similarly, at the second-order, we have\ea{{\label{con_sol_2}}
    \alpha_2 &= \alpha_1^2 + \frac{1}{2H} \rd^{-2} \rd_i  \Gamma_i
    \,,\\
    \beta_2  &= \frac{a^2}{4H}  \rd^{-2}\left\{ \lrsb{ -P_{,\ab{IJ}} \dop_0^I \dop_0^J -   P_{,\ab{IJ}\ab{KL}}  \dop_0^I
        \dop_0^J \dop_0^K \dop_0^L
     +6H^2 }
        (3\alpha_1^2 - 2\alpha_2)  \right.\\
        &\qquad\qquad\qquad \left. + 2\, \Omega - \frac{1}{a^4} \lrsb{ \rd_i\rd_j\beta_1\rd_i\rd_j\beta_1 - \lrp{ \rd^2\beta_1 }^2 }  + \frac{8H\alpha_1 }{a^2}
        \rd^2\beta_1 \right\} \\
        \theta_{2i} &= 2a^2 \lrp{ \frac{\rd_i
        \rd_j}{\rd^4}
        - \frac{\delta_{ij}}{\rd^2} } \Gamma_j \,,\\
} with
    \ea{
        \Gamma_i & \equiv -   \frac{\rd_j \alpha_1}{a^2} \lrsb{ \rd_i \rd_j \beta_1 - \rd^2 \beta_1 \delta_{ij}} \\
        &\qquad \qquad + \lrsb{ \lrp{ P_{,\ab{IJ}\ab{KL}} \lrp{\dop_0^{(K}
\doq^{L)}  - \alpha_1\, \dop_0^K \dop_0^L } + P_{,\ab{IJ}K}Q^K
-\alpha_1 P_{,\ab{IJ}} } \dop_0^I + P_{,\ab{IJ}} \doq^I }
        \rd_iQ^J \,, }
and
\ea{
        \Omega &\equiv \lrp{ P_{,\ab{IJ}} - P_{,\ab{IJ}\ab{KL}} \dop_0^K \dop_0^L }
\lrcb{ -\frac{1}{2a^2} \rd_i Q^I \rd_i Q^J  + \frac{1}{2} \lrsb{
\lrp{ \doq^I \doq^J - 2\, \dop_0^{(I} N^i_{1} \rd_i Q^{J)}  } -
4\alpha_1 \, \dop_0^{(I} \doq^{J)} }
        } \\
        &\qquad +  \frac{1}{2}  \lrp{ P_{,\ab{KL}\ab{MN}}  - \dop_0^I\dop_0^J P_{,\ab{IJ}\ab{KL}\ab{MN} } } \lrp{  \dop_0^{(K} \doq^{L)}  - \alpha_1\, \dop_0^K \dop_0^L } \lrp{ \dop_0^{(M} \doq^{N)}  - \alpha_1\, \dop_0^M \dop_0^N } \\
        &\qquad + \lrp{ P_{,\ab{KL}M} -\dop_0^I\dop_0^J P_{,\ab{IJ}\ab{KL} M} } Q^M \lrp{  \dop_0^{(K} \doq^{L)}  - \alpha_1\, \dop_0^K \dop_0^L } \\
        &\qquad + \frac{1}{2}\lrp{
        P_{,KL} - \dop_0^I\dop_0^J P_{,\ab{IJ}KL} }
        Q^KQ^L + 2\alpha_1  \lrsb{ P_{,\ab{IJ}\ab{KL}}  \lrp{ \dop_0^{(K} \doq^{L)}  - \alpha_1\, \dop_0^K \dop_0^L  } + P_{,\ab{IJ}K}Q^K } \dop_0^I\dop_0^J    \\
        &\qquad\qquad   - \lrsb{ P_{,\ab{IJ}\ab{KL}}  \lrp{ \dop_0^{(K} \doq^{L)}  - \alpha_1\, \dop_0^K \dop_0^L  } + P_{,\ab{IJ}K}Q^K -2\alpha_1 P_{,\ab{IJ}0}
    } 2\, \dop_0^{(I} \doq^{J)} \\
    &\qquad - P_{,\ab{IJ}} \lrp{ \doq^I \doq^J - 2\, \dop_0^{(I} N^i_{1} \rd_i Q^{J)}
    } \,.
    }
It is useful to compare our above results with previously known
results
\cite{Seery:2006vu,Seery:2006js,Langlois:2008qf,Arroja:2008yy,Arroja:2008ga},
which the above results reduce to in various limits.

\section{Linear Perturbations}

\subsection{Instantaneous Adiabatic/entropy Modes Decomposition}

In multi-field inflation models, it is convenient to decompose
perturbations into instantaneous adiabatic and entropy perturbations
\cite{Gordon:2000hv,Bassett:2005xm}. The ``adiabatic direction"
corresponds to the direction of the ``background inflaton velocity"
    \eq{
        e^I_1 \equiv \frac{\dop^I}{ \sqrt{ P_{,\ab{JK}} \dop^J\dop^K
        }} \equiv \frac{\dop^I}{\dot{\sigma}}\,,
    }
where we define $\dot{\sigma}$
    \eq{{\label{dot_sigma}}
        \dot{\sigma} \equiv \sqrt{ P_{,\ab{JK}} \dop^J\dop^K} \,,
    }
which is the generalization of the background inflaton velocity.
Actually $\dot{\sigma}$ is essentially a short notation and has
nothing to do with any concrete field. Note that $\dot{\sigma}$ is
related to the slow-roll parameter $\epsilon$ as $\dot{\sigma}^2 =
2H^2 \epsilon$.

We introduce $(\mathcal{N}-1)$ basis $e^I_n$,
($n=2,\cdots,\mathcal{N}$) which are orthogonal with $e^I_1$ and
also with each other. The orthogonal condition can be defined as
    \eq{{\label{orthogonal}}
        P_{,\ab{IJ}} e^I_m e^J_n \equiv \delta_{mn} \,.
    }
Thus the scalar-field perturbation $Q^I$ can be decomposed into
instantaneous adiabatic/entropy basis:
    \eq{
        Q^I \equiv e^I_m Q^m \,,\qquad m=1,\cdots \mathcal{N} \,.
    }

Up to now our discussion is rather general, without further
restriction on the structure of $P(X^{IJ},\phi^I)$. In this work, we
consider a general class of two-field model, with the Lagrangian for
the scalar fields as the following form{\footnote{ This form of
Lagrangian if motivated from that, for multi-field $k$-inflation
models \cite{Langlois:2008mn,Gao:2008dt}, the Lagrangian is simply
$P(X,\phi^I)$, In \cite{Arroja:2008yy} a special form of Lagrangian
$ \tilde{P}(\tilde{Y},\phi^I) $ with $\tilde{Y} \equiv X +
\frac{b(\phi^I)}{2} \lrp{ X^2 - X_{IJ}X^{IJ} }$ was chosen in the
investigation of bispectrua in two-field models, which is motivated
by multi-field DBI action. In this work, we use the more general
form of Lagrangian (\ref{specific_model}).}}:
    \eq{{\label{specific_model}}
        P(X^{IJ},\phi^I) = P(X,Y,\phi^I) \,,
    }
with $X\equiv  X^I_I =  G_{IJ}X^{IJ}$ and $Y\equiv X^I_J X^J_I $.
This form of Lagrangian not only is the most general Lagrangian for
two-field models and thus deserves detailed investigations, but also
can make our discussions on the non-Gaussianities in two-field
models in a more general background. For two-field case, higher
contractions among $X_{IJ}$ can be expressed in terms of $X$ and
$Y$, e.g.,
    \[
    \begin{aligned}
        X^I_J X^J_K X^K_I &= -\frac{X^{3}}{2}+\frac{3}{2}XY \,,\\
        X^I_J X^J_K X^K_L X^L_I &= -\frac{1}{2}X^{4}+X^{2}Y+\frac{1}{2}Y^{2}
        \,,
    \end{aligned}
    \]
etc. The model (\ref{specific_model}) includes multi-field
$k$-inflation and two-field DBI model as special cases. For example,
in multi-DBI model the Lagrangian is $P = - \frac{1}{f(\phi^I)}
\lrp{ \sqrt{ \mathcal{D} }-1} -
        V(\phi^I)$
with
    \ea{
        \mathcal{D} &\equiv \det \lrp{ G^I_J - 2f X^I_J } \\
        &= 1 -2f G_{IJ}X^{IJ} + 4f^2 X^{[I}_I X^{J]}_J - 8f^3
        X^{[I}_I X^J_J X^{K]}_K + 16 f^4 X^{[I}_I X^J_J X^K_K
        X^{L]}_L \,.
    }
This expression for determinant $\mathcal{D}$ is general. In this
work, we focus on two-field case, thus the last two terms exactly
vanish, leaving as effectively $\mathcal{D} \equiv 1 -2f
G_{IJ}X^{IJ} + 4f^2 X^{[I}_I
        X^{J]}_J $. In terms of (\ref{specific_model}), this is $\mathcal{D} = 1-2fX+2f^2\lrp{X^2-Y}$.

At this point, it is convenient to introduce two parameters:
    \ea{{\label{def_lambda_Pi}}
        \lambda &\equiv X^{2}P_{,XX}+\frac{2}{3}X^{3}P_{,XXX}+2\left(YP_{Y}+6Y^{2}P_{,YY}+\frac{8}{3}Y^{3}P_{,YYY}\right) \\
        &\qquad +4\left(X^{2}YP_{,XXY}+2XYP_{,XY}+2XY^{2}P_{,XYY}\right) \,,\\
        \Pi &\equiv X^{3}P_{,XXX}+\frac{2}{5}X^{4}P_{,XXXX}+\frac{4}{5}\left(21Y^{2}P_{,YY}+34Y^{3}P_{,YYY}+8Y^{4}P_{,YYYY}\right)\\
        &\qquad +\frac{16}{5}X^{3}YP_{,XXXY}+\frac{54}{5}X^{2}YP_{,XXY}+\frac{48}{5}X^{2}Y^{2}P_{,XXYY}+6XYP_{,XY} \\
        &\qquad +\frac{156}{5}XY^{2}P_{,XYY}+\frac{64}{5}XY^{3}P_{,XYYY}
        \,,
    }
where all quantities are background values, and we have used
$Y=X^2$. As we will see later, although the $X$,$Y$-dependences of
$P(X,Y,\phi^I)$ in general can be complicated, the non-linear
structures of $P$ affect the trispectra through the above specific
combinations of derivatives of $P$.

\subsection{Quantization and the Power Spectra}

After instantaneous adiabatic/entropy modes decomposition, at the
leading-order, the second-order action for the perturbations takes
the form{\footnote{In (\ref{S2_Q}) we neglect the mass-square terms
as $\mathcal{M}_{mn}Q^m Q^n$ and the friction terms such as $\sim
\dot{Q}_m Q_n$. In general these terms may become important,
especially they may cause non-vanishing cross-correlations between
adiabatic mode and entropy mode around horizon-crossing. See e.g.
\cite{Arroja:2008yy,Langlois:2008mn,Langlois:2008qf} for details.}}
 \eq{{\label{S2_Q}}
    S_2^{\textrm{(main)}} = \int dt d^3x\, a^3 \lrp{ \frac{1}{2} \mathcal{K}_{mn} \doq_m \doq_n - \frac{1}{2a^2} \delta_{mn} \rd_i Q_m \rd_i
        Q_n} \,,\\
} with \ea{{\label{K_mn}}
    \mathcal{K}_{mn} &\equiv \delta_{mn} + \lrp{P_{,\ab{MN}} \dop^M
        \dop^N} P_{,\ab{IK} \ab{JL}} e^I_1 e^K_n e^J_1 e^L_m \,,\\
        &= \delta_{mn} + \lrp{ \frac{1}{\ca^2}-1 }
        \delta_{1m} \delta_{1n} + \lrp{ \frac{1}{\ce^2} - 1 }
        \lrp{\delta_{mn}- \delta_{1m} \delta_{1n}
        } \,,
        }
where we introduce{\footnote{We use $\ca$ and $\ce$ rather than
$c_{\sigma}$ and $c_s$ in order to avoid possible confusion, since
in the literatures $c_s$ has special meaning, i.e. the speed of
sound of perturbation in single-field models.}}
    \ea{{\label{def_ca_ce}}
        \ca^2 &\equiv \frac{P_{,X}+2XP_{,Y}}{P_{,X}+2X\left(P_{,XX}+4XP_{,XY}+3P_{,Y}+4X^{2}P_{,YY}\right)} \,,\\
        \ce^2 &\equiv \frac{P_{,X}}{P_{,X}+2XP_{,Y}} \,,
    }
which are the propagation speeds of adiabatic perturbation and
entropy perturbation respectively. It is useful to note that
$\mathcal{K}_{mn}$ is diagonal, $\mathcal{K}_{11} = 1/\ca^2$,
$\mathcal{K}_{22} = 1/\ce^2$ and $\mathcal{K}_{12} =
\mathcal{K}_{21} = 0$, as a sequence of adiabatic/entropy
decomposition. $\ca \neq \ce$ is a generic feature in multi-field
models, this can also be seen explicitly from the definitions in
(\ref{def_ca_ce}), the speed of sound for adiabatic mode and entropy
mode(s) have different dependence of $P$-derivatives\footnote{This
fact was first point out apparently in
\cite{Easson:2007dh,Huang:2007hh} in the investigation of brane
inflation models. See also
\cite{Arroja:2008yy,Langlois:2008qf,Cai:2008if,Cai:2009hw,Ji:2009yw,Gao:2008dt}
for extensive investigations on general multi-field models with
different $\ca$ and $\ce$.}.

Obviously, $\qsg$ and $\qs$ themselves are not properly normalized
 for canonical quantization. We may introduce new variables
    \eq{
        \tilde{Q}_{\sigma} \equiv  \frac{a}{\ca} \qsg \,,\qquad\qquad \tilde{Q}_{s}
        \equiv \frac{a}{\ce} \qs \,,
    }
which are canonically normalized variables, since after changing
into comoving time defined by $dt=a d\eta$, the quadratic action
takes the form
    \eq{{\label{S2_tilde_Q}}
        S_2 = \int d\eta d^3x\,\frac{1}{2} \lrsb{ \tilde{Q}'^2_{\sigma} + \lrp{ \mathcal{H}^2 + \mathcal{H}' } \tilde{Q}_{\sigma}^2 - \ca^2 (\rd \tilde{Q}_{\sigma})^2 + \tilde{Q}'^2_{s} + \lrp{ \mathcal{H}^2 + \mathcal{H}' } \tilde{Q}_{s}^2 - \ce^2 (\rd \tilde{Q}_{s})^2   }\,.
    }
The action (\ref{S2_Q}) or (\ref{S2_tilde_Q}) describes a free
theory, which is easy to quantize. In canonical quantization, we
write
    \eq{
        \tilde{Q}_{\sigma}(\vk,\eta) \equiv a_{\vk} \tilde{u}_k(\eta) + a^{\dag}_{-\vk}
        \tilde{u}^{\ast}_k(\eta) \,, \qquad\qquad \tilde{Q}_s(\vk,\eta) \equiv a_{\vk} \tilde{v}_k(\eta) + a^{\dag}_{-\vk}
        \tilde{v}^{\ast}_k(\eta) \,,
    }
where $\tilde{u}_k(\eta)$ and $\tilde{v}_k(\eta)$ are the mode
functions, which satisfy the corresponding classical equations of
motion
    \ea{{\label{mode_function_tilde}}
        \tilde{u}''_k + \lrsb{ \ca^2 k^2 - ( \mathcal{H}^2 + \mathcal{H}' ) } \tilde{u}_k =0   \,,\qquad\qquad \tilde{v}''_k + \lrsb{ \ce^2 k^2 - ( \mathcal{H}^2 + \mathcal{H}' )
}  \tilde{v}_k  =0  \,,
    }
and can be easily solved in de Sitter approximation ($a(\eta) =
-\frac{1}{H\eta}$):
    \ea{
        \tilde{u}_k(\eta) = \frac{1}{\sqrt{ 2 \ca k}} e^{-i\ca k \eta} \lrp{ 1 - \frac{i}{\ca k \eta} }  \,,\qquad\qquad  \tilde{v}_k(\eta)
        = \frac{1}{\sqrt{ 2 \ce k}} e^{-i\ce k \eta} \lrp{ 1 - \frac{i}{\ce k \eta} }
        \,.
    }
Note that the mode functions are chosen so that when the modes are
deep in the sound horizon, or equivalently $\eta\rightarrow
-\infty$, they behave as free harmonic oscillators in Minkowski
spacetime, i.e.
    \[
        \tilde{u}_k(\eta) \xrightarrow[]{\eta \rightarrow -\infty}
        \frac{1}{\sqrt{2 \ca k}} e^{-i \ca k\eta}
        \,,\qquad\qquad \tilde{v}_k(\eta) \xrightarrow[]{\eta \rightarrow
        -\infty} \frac{1}{\sqrt{2 \ce k}} e^{-i \ce k\eta} \,.
    \]
Moreover, $\tilde{u}_k(\eta)$ and  $\tilde{v}_k(\eta)$ are
normalized with Wronskian
    \eq{
        \tilde{u}_k(\eta)\tilde{u}'^{\ast}_k(\eta) -
        \tilde{u}^{\ast}_k(\eta)\tilde{u}'_k(\eta) \equiv \tilde{v}_k(\eta)\tilde{v}'^{\ast}_k(\eta) -
        \tilde{v}^{\ast}_k(\eta)\tilde{v}'_k(\eta) \equiv i \,,
    }
which is the condition for canonical quantization.

Finally, what we are interested in are the tree-level two-point
functions for $\qsg$ and $\qs$, defined as
    \ea{
        \lrab{ \qsg(\vk_1,\eta_{1}) \qsg(\vk_2,\eta_{2}) } = (2\pi)^3 \delta^2 (\vk_1 + \vk_2)
        G_{k_1}(\eta_1,\eta_2) \,,\\
        \lrab{ \qs(\vk_1,\eta_{1}) \qs(\vk_2,\eta_{2}) } = (2\pi)^3 \delta^2 (\vk_1 + \vk_2)
        F_{k_1}(\eta_1,\eta_2) \,,\\
    }
with
    \eq{
        G_k(\eta_1,\eta_2) \equiv u_k(\eta_1) u^{\ast}_k(\eta_2)
        \,,\qquad \qquad F_k(\eta_1,\eta_2) \equiv v_k(\eta_1)
        v^{\ast}_k(\eta_2) \,,
    }
where $u_k(\eta)$ and $v_k(\eta)$ are the mode functions for
adiabatic perturbation and entropy perturbation respectively:
    \ea{
        u_k(\eta) &= \frac{i\, H}{ \sqrt{ 2\ca k^3 }} \lrp{ 1+ i\ca k \eta } e^{-i \ca k \eta}  \,,\\
        v_k(\eta) &=  \frac{i\, H}{ \sqrt{ 2\ce k^3 }} \lrp{ 1+ i\ce k \eta } e^{-i \ce k \eta} \,.
    }

The so-called ``power spectra" for adiabatic perturbation and
entropy perturbation are defined as $P_{\sigma}(k) \equiv
G_k(\eta_{\ast},\eta_{\ast})$ and $P_{s}(k) \equiv
F_k(\eta_{\ast},\eta_{\ast})$, where $\eta_{\ast}$ can be chosen as
the time when the modes cross the sound-horizon, i.e. at $\ca
k\equiv aH$ for adiabatic mode and $\ce k \equiv aH$ for entropy
mode(s){\footnote{In general multi-field models, adiabatic mode and
entropy modes(s) with the same comoving wavenumber $k$ exit the
sound-horizon at different time, due to their different speeds of
sound, $\ca \neq \ce$. This phenomena may cause subtle problems,
such as the problem of decoherences. In this work, we neglect these
problems.}}. We have
    \eq{
        P_{\sigma\ast}(k) = \frac{H^2}{2\ca k^3}\,,\qquad\qquad P_{s\ast}(k) =
        \frac{H^2}{2\ce k^3} \,.
    }

In the so-called comoving gauge, the perturbation $\qsg$ is directly
related to the three-dimensional curvature of the constant time
space-like slices. This gives the gauge-invariant quantity referred
to as the ``comoving curvature perturbation":
    \eq{
        \mathcal{R} \equiv \frac{H}{\dot{\sigma}} \qsg \,,
    }
where $\dot{\sigma}$ is defined in (\ref{dot_sigma}). The entropy
perturbation $\qs$ is automatically gauge-invariant by construction.
It is also convenient to introduce a renormalized ``isocurvature
perturbation" defined as
    \eq{
        \mathcal{S} \equiv \frac{H}{\dot{\sigma}} \qs \,.
    }

In the cosmological context, it is also convenient to define the
dimensionless power spectra for comoving curvature perturbation and
isocurvature perturbation respectively:
    \ea{{\label{dimless_spec}}
        \mathcal{P}_{\mathcal{R}\ast} &= \frac{H^2}{\dot{\sigma}^2} \mathcal{P}_{\sigma\ast} \equiv \frac{H^2}{\dot{\sigma}^2} \frac{k^3}{2\pi^2} P_{\sigma\ast}(k)
        = \frac{1}{2\epsilon\ca} \lrp{ \frac{H}{2\pi} }^2\,,\\
        \mathcal{P}_{\mathcal{S}\ast} &= \frac{H^2}{\dot{\sigma}^2} \mathcal{P}_{s\ast} \equiv \frac{H^2}{\dot{\sigma}^2} \frac{k^3}{2\pi^2}  P_{s\ast}(k)
        = \frac{1}{2\epsilon\ce} \lrp{ \frac{H}{2\pi} }^2 \,.
    }
In the above results, all quantities are evaluated around the
sound-horizon crossing. $\mathcal{P}_{\mathcal{R}}$ in
(\ref{dimless_spec}) recovers the well-known result for single-field
models \cite{ArmendarizPicon:1999rj,Garriga:1999vw,Chen:2006nt}. In
the case when $\ca = \ce$, the above results reduce to those in
multi-field DBI model which has been investigated in
\cite{Langlois:2008qf,Langlois:2008wt,Arroja:2008yy}.

\subsection{Superhorizon Evolution}

Actually, the inflaton fields perturbation, or $\qsg$ and $\qs$
themselves are not directly observable. What we are interested in is
the curvature perturbation. In single-field inflation models, this
has no particular difficulties since the comoving curvature
perturbation $\mathcal{R}$ is conserved on superhorizon scales
\cite{Lyth:2004gb,Langlois:2005ii,Langlois:2005qp}. Thus, it is
sufficient to evaluate the correlation functions for $\zeta$, i.e.
the curvature perturbation on uniform density hypersurfaces which
coincides with $-\mathcal{R}$ on superhorizon scales, at the time of
horizon-crossing.

However, in contrast with the single-field models, in multi-field
models, the curvature perturbation in general evolves after the
horizon-crossing \cite{Starobinsky:1994mh} (see also
\cite{Lalak:2007vi}). This is due to the fact that, in superhorizon
scales adiabatic perturbation can be sourced by entropy
perturbation(s) and there is a transfer between adiabatic/entropy
modes. This can be clearly seen if we take the time derivative of
the curvature perturbation \cite{Gordon:2000hv} (see
\cite{Langlois:2008mn} for a recent investigation), in our model it
is
    \eq{
        \dot{\mathcal{R}} \equiv \frac{H}{\dot{H}} \frac{\ca^2 k^2}{a^2} \Psi + \tilde{\xi} \mathcal{S}
        \,,
    }
where
    \eq{{\label{tilde_xi}}
    \tilde{\xi} \equiv  \frac{(1+\ca^2) \tilde{P}_{,s} - \ca^2 \dot{\sigma}^2 \tilde{P}_{,Ys} }{\dot{\sigma}\ca \tilde{P}_{,Y} }
}
 Due to the presence of isocurvature perturbation
$\mathcal{S}$, even on superhorizon scales $|\ca k/a|\ll 1$,
$\dot{\mathcal{R}} \approx \tilde{\xi} \mathcal{S} \neq 0$, that is
$\mathcal{R}$ will evolve. In general, on superhorizon scales, the
evolution of curvature/isocurvature perturbations can be
approximately described by
    \eq{
        \dot{\mathcal{R}} \approx \alpha H
        \mathcal{S}\,,\qquad\qquad \dot{\mathcal{S}} \approx \beta H
        \mathcal{S} \,.
    }
 The above equations have formal solutions
    \eq{
        \lrp{\begin{array}{c}
               \mathcal{R} \\
               \mathcal{S}
             \end{array}
         } = \lrp{ \begin{array}{cc}
                     1 & T_{\mathcal{R}\mathcal{S}} \\
                     0 & T_{\mathcal{S}\mathcal{S}}
                   \end{array}
          } \lrp{\begin{array}{c}
               \mathcal{R} \\
               \mathcal{S}
             \end{array}
         }_{\ast} \,,
    }
with
    \eq{
        T_{\mathcal{S}\mathcal{S}}(t,t_{\ast}) = e^{\int_{t_{\ast}}^t dt'\, \beta(t')H(t')
        }\,,\qquad T_{\mathcal{R}\mathcal{S}}(t,t_{\ast}) = \int_{t_{\ast}}^t
        dt'\, \alpha(t') T_{\mathcal{S}\mathcal{S}}(t',t_{\ast})
        H(t') \,,
    }
where $t_{\ast}$ is the time of horizon-crossing and $t$ is some
later time. Thus on superhorizon scales, the (time-dependent) power
spectra for the curvature perturbation, isocurvature perturbation
and also the cross-correlation between the two can be formally
expressed as
    \ea{{\label{spectra_ls}}
        \mathcal{P}_{\mathcal{R}}(t) &= \mathcal{P}_{\mathcal{R}\ast} +
        T^2_{\mathcal{R}\mathcal{S}}(t,t_{\ast})
        \mathcal{P}_{\mathcal{S}\ast} \,,\\
        \mathcal{P}_{\mathcal{S}}(t) &=
        T^2_{\mathcal{S}\mathcal{S}}(t,t_{\ast})
        \mathcal{P}_{\mathcal{S}\ast} \,,\\
        \mathcal{C}_{\mathcal{R}\mathcal{S}}(t) &\equiv \lrab{ \mathcal{R}\mathcal{S} }=
        T_{\mathcal{R}\mathcal{S}}(t,t_{\ast})T_{\mathcal{S}\mathcal{S}}(t,t_{\ast})
        \mathcal{P}_{\mathcal{S}\ast} \,.\\
    }

\section{Non-linear Perturbations}

In this section, we derive the fourth-order action for the
perturbations and the fourth-order interaction Hamiltonian (in the
interaction picture). In the next section, we evaluate the
four-point correlation functions for the perturbations.

\subsection{Fourth-order Perturbation Action}

From (\ref{action_einstein}), Taylor expansion gives the
fourth-order perturbation action from the gravity sector
 \ea{
    S^{g}_4 &= \int dtd^3x\, \frac{a^3}{2} \left\{ -6H^2 \lrp{ \alpha _1^4- 3 \alpha _1^2 \alpha _2+\alpha_2^2 } + \lrp{ -\alpha _1^3 + 2 \alpha _1 \alpha _2} \lrp{  \frac{4H}{a^2}  \rd^2 \beta_1 } \right.\\
    &\qquad\qquad +  \lrp{ \alpha_1^2 -\alpha _2} \lrp{  \frac{4H}{a^2}  \rd^2 \beta_2 + \frac{1}{a^4} \lrsb{ \rd_i \rd_j \beta_1 \rd_i \rd_j \beta_1  - \lrp{ \rd^2\beta_1 }^2 } } \\
    &\qquad\qquad \left. - \frac{2\alpha_1}{a^4} \lrsb{
    (\rd_i\rd_j \beta_1) \lrp{ \rd_i\rd_j \beta_2 + \rd_{(i} \theta_{2j)}
    } - \rd^2 \beta_1 \rd^2 \beta_2
    } +  \frac{1}{a^4} \lrsb{ 2\, \rd_i\rd_j \beta_2\,  \rd_i\theta_{2j} + \rd_i\theta_{2j} \rd_{(i} \theta_{2j)}  }
    \right\}  \,.
} The scalar field sector is more complicated,
    \eq{
        S^{\phi}_4 \equiv  \int dtd^3x\, a^3 \lrp{ P_4 + \alpha_1 P_3 + \alpha_2
        P_2 }\,,
    }
where $P_n$'s are the corresponding parts in the expansion of $P$
which are order $\mathcal{O}(Q^n)$ respectively, which we do not
wrtie explicitly here for clarity, and can be found in Appendix
\ref{app_stacks}.

\subsubsection{Fourth-order Action at the Leading-order}

In this work, we focus on the dominant contributions to the
non-Gaussianities from the trispectra for $\ca \ll 1$ and $\ce \ll
1$. In the case of bispectrum, in this limit, contributions from the
gravity sector $S^g_3$ and the metric perturbations themselves
$\alpha_1$, $\beta_1$ can be neglected in the leading order
\cite{Langlois:2008qf}. However, this is no longer the case for the
fourth-order calculation. The orders of various quantities can be
read from (\ref{con_sol_1})-(\ref{con_sol_2}) and are summarized in
Tab.\ref{tab_order}.
\begin{table}[h]
\centering
    \begin{tabular}{c|c|c|c|c|c}
      \hline
      $\alpha_1$ & $\beta_1$ & $\theta_{1i}$ & $\alpha_2$ & $\beta_2$ & $\theta_{2i}$ \\
      \hline
      $\mathcal{O}(\sqrt{\epsilon} )$ & $\mathcal{O}( \sqrt{\epsilon}/c_s^2 )$ & - & $\mathcal{O}(1/c_s^2)$ & $\mathcal{O}(1/c_s^2)$ & $\mathcal{O}(1/c_s^2)$ \\
      \hline
    \end{tabular}
    \caption{Summary of the order of constraints. $c_s$ denotes both $\ca$ and $\ce$.}
    \label{tab_order}
    \end{table}

Moreover, in the slow-roll limit, the fourth-order action from the
gravity sector can be approximately neglected, since $
\frac{\mathcal{L}^g_4}{ \mathcal{L}^\phi_4} \sim \epsilon$. Thus, at
leading-order in slow-roll, the fourth-order perturbation action
reads \ea{{\label{4th_action_IJ}}
    S_4^{\textrm{(main)}} &= \int dtd^3x\, a^3 \left[ \frac{1}{2} P_{,\ab{IJ}\ab{KL}} X^{IJ}_2 X^{KL}_2 + \frac{1}{2} P_{,\ab{IJ} \ab{KL}\ab{MN}} X^{IJ}_2 X^{KL}_1 X^{MN}_1 \right. \\
                    &\qquad\qquad\qquad\qquad \left. +\frac{1}{24} P_{,\ab{IJ} \ab{KL} \ab{MN} \ab{PQ} } X^{IJ}_1 X^{KL}_1 X^{MN}_1 X^{PQ}_1
                    \right] \\
                    &\simeq \int dtd^3x\, a^3 \left\{ \Gamma_{IJKL}\, \doq^{I}  \doq^{J}  \doq^{K}
                    \doq^{L} - \Theta_{IJKL}\, \frac{1}{4a^2} \doq^I \doq^J\, \rd_i Q^K \rd_i
                    Q^L  \right.\\
                    &\qquad\qquad\qquad\qquad \left. + P_{,\ab{IJ}\ab{KL}} \frac{1}{8a^4} \rd_i Q^I \rd_i Q^J \, \rd_j Q^K \rd_j Q^L
                    \right\} \,,
    }
where
    \ea{
        \Gamma_{IJKL} &\equiv \frac{1}{8}  P_{,\ab{IJ}\ab{KL}} +  \frac{1}{4} P_{,\ab{IJ} \ab{MK}\ab{NL}} \dop_0^{M} \dop_0^{N} +\frac{1}{24} P_{,\ab{MI} \ab{NJ} \ab{PK} \ab{QL} }
                    \dop_0^{M} \dop_0^{N} \dop_0^{P} \dop_0^{Q}
                    \,,\\
        \Theta_{IJKL} &\equiv P_{,\ab{IJ}\ab{KL}}  + P_{,\ab{KL} \ab{MI}\ab{NJ}} \dop_0^{M}
        \dop_0^{N} \,,
    }
for simplicity. The explicit expression for $X^{IJ}_2$ can be found
in Appendix \ref{app_stacks}.

After adiabatic/entropy decomposition, the fourth-order action
(\ref{4th_action_IJ}) can be written as
    \ea{{\label{4th_action__mn}}
        S_4^{\textrm{(main)}} \simeq \int dt d^3x\,a^3 \lrsb{ \Gamma_{mnpq}\, \doq_m  \doq_n \doq_p \doq_q - \frac{1}{4a^2} \Theta_{mnpq}\, \doq_m \doq_n \rd_i Q_p \rd_i Q_q + \frac{1}{8a^4}\Omega_{mnpq}\, \rd_i Q_m \rd_i Q_n \rd_j Q_p \rd_j Q_q
        } \,,
    }
where
    \ea{
        \Gamma_{mnpq} &= \frac{1}{8}  P_{,\ab{IJ}\ab{KL}}\, e^I_m e^J_n e^K_p e^L_q +  \frac{1}{4}\lrp{ P_{,\ab{PQ}} \dop^P \dop^Q } P_{,\ab{IJ} \ab{MK}\ab{NL}}  e^M_1 e^N_1 \, e^I_m e^J_n e^K_p e^L_q \\
        &\qquad\qquad +\frac{1}{24} \lrp{ P_{,\ab{UV}} \dop^U \dop^V }^2 P_{,\ab{MI} \ab{NJ} \ab{PK} \ab{QL} }
                    e^M_1 e^N_1 e^P_1 e^Q_1 \, e^I_m e^J_n e^K_p
                    e^L_q \,,\\
        \Theta_{mnpq} &=  P_{,\ab{IJ}\ab{KL}} \, e^I_m e^J_n e^K_p e^L_q + \lrp{ P_{,\ab{PQ}} \dop^P \dop^Q } P_{,\ab{KL} \ab{MI}\ab{NJ}}
        e^M_1 e^N_1 e^I_m e^J_n e^K_p e^L_q \,,\\
        \Omega_{mnpq} &= P_{,\ab{IJ}\ab{KL}}\, e^I_m e^J_n e^K_p
        e^L_q \,.
    }

\subsection{Interacting Hamiltonian}

In the operator formalism of quantization, interaction Hamiltonian
is needed. The interactions of cosmological perturbation are in
general contain time derivatives, which are different form ordinary
field theory where interactions are local products of fields. Thus,
in order to get the corresponding Hamiltonian, we should use its
definition
    \eq{
        \mathcal{H} \equiv \pi_I \doq^I -
        \mathcal{L} \,,
    }
where $\mathcal{L}$ is the Lagrangian containing 2nd, 3rd and
4th-order terms. The 2nd-order part is given in (\ref{S2_Q}). The
3rd-order part has been derived in \cite{Arroja:2008yy}:
    \ea{{\label{action_2nd_3rd}}
        S_3^{\textrm{(main)}} &= \int dt d^3x\, a^3 \lrp{ \frac{1}{2} \Xi_{mnl} \doq_m \doq_n \doq_l  - \frac{1}{2a^2} \Upsilon_{mnl}\, \doq_m \rd_iQ_n \rd_iQ_l } \,,
    }
with
    \ea{
        \Xi_{mnl} &\equiv \sqrt{ P_{,\ab{MN}} \dop^M\dop^N
        } \lrsb{ P_{,\ab{IK} \ab{JL}}  e^{I}_1 e^{K}_{m} e^J_{n} e^L_{\ell} +\frac{1}{3} \lrp{ P_{,\ab{MN}} \dop^M\dop^N } P_{,\ab{IK} \ab{JL} \ab{PQ}
        } e^{I}_1 e^{K}_m e^{J}_1 e^{L}_n e^{P}_1 e^{Q}_{l} }
        \,,\\
        \Upsilon_{mnl} &\equiv \sqrt{P_{,\ab{MN}}
        \dop^M\dop^N} \, P_{,\ab{IK} \ab{JL}}\,  e^{I}_1 e^{K}_m e^{J}_n e^L_l \,.
    }

From (\ref{S2_Q}), (\ref{4th_action__mn}) and
(\ref{action_2nd_3rd}), through a straightforward but rather tedious
calculation, we get
    \ea{{\label{H4_final}}
        \mathcal{H}_4 &= \doq_m \doq_n \doq_p \doq_q \lrsb{ \frac{9}{8} \mathcal{K}^{-1}_{rs}\Xi_{rmn} \, \Xi_{spq}  - \Gamma_{mnpq}
        } + \frac{1}{a^2} \doq_m \doq_n \rd_i Q_p \rd_i Q_q \lrp{ \frac{1}{4}
        \Theta_{mnpq}- \frac{3}{4} \mathcal{K}^{-1}_{rs}\Upsilon_{rpq}\, \Xi_{smn}
        } \\
        &\qquad + \frac{1}{a^4} \rd_i Q_m \rd_i Q_n \rd_j Q_p \rd_j Q_q  \lrp{ \frac{1}{8}
        \mathcal{K}^{-1}_{rs} \Upsilon_{rmn} \,\Upsilon_{spq}  -  \frac{1}{8}
        \Omega_{mnpq} } \,.
    }
See Appendix \ref{appsec_Hamiltonian} for detailed derivations.

After a straightforward calculation and changing into comoving time
$\eta$ defined by $dt = a\, d\eta$, we get the leading-order
4th-order (interaction picture) interaction Hamiltonian in terms of
$\qsg$ and $\qs$, for the model (\ref{specific_model}):
    \eq{{\label{H4_struc}}
        H_4(\eta) \equiv \int d\eta d^3x\, \lrp{ \mathcal{H}_4^{\sigma} + \mathcal{H}_4^{\sigma} + \mathcal{H}_4^{c}
        } \,,
    }
with
 \begin{align}
    \mathcal{H}^{\sigma}_4 &= \Gamma_{\sigma}\, \qsg'^4 + \Theta_{\sigma}\, \qsg'^2 \lrp{ \rd \qsg}^2 +  \Omega_{\sigma}\, \lrp{ \rd_i \qsg \rd_i \qsg }^2
    \,, \label{H4_sigma}\\
    \mathcal{H}_4^{s} &= \Gamma_s\, \qs'^4 +  \Theta_{s}\, \qs'^2 \lrp{ \rd
    \qs}^2  + \Omega_s\, \lrp{ \rd_i \qs \rd_i \qs
    }^2  \,, \label{H4_s}\\
    \mathcal{H}_4^c &= \Gamma_c \, \qsg'^2 \qs'^2  +  \Theta_{\sigma s}\, \qsg'^2 \lrp{ \rd
    \qs}^2 + \Theta_{s\sigma } \, \qs'^2 \lrp{ \rd \qsg}^2 +  \Theta_{c} \, \qsg' \qs'  \lrp{ \rd_i \qsg \rd_i
    \qs} \nonumber \\
    &\qquad +
    \Omega_{\sigma s}\,  \lrp{\rd\qsg}^2 \lrp{\rd\qs}^2 +
    \Omega_{c}\,
    \lrp{ \rd_i \qsg \rd_i \qs }^2  \,, \label{H4_cross}
    \end{align}
where the various coefficients can be found in Appendix
\ref{appsec_H4_coeff}. It can be seen directly from
(\ref{H4_struc})-(\ref{H4_cross}) that there are three types of
four-point interaction vertices, one involving four temporal
derivatives, one involving four spatial derivatives and one
involving two temporal and two spatial derivatives. This is similar
to the case in general single field inflation. Actually as we will
see below, the four-point functions can be grouped into three types
which correspond to three fundamental momentum-dependent shape
functions. Moreover, there are 4$\qsg$, 4$\qs$ and 2$\qsg$2$\qs$
interactions. Thus the corresponding non-vanishing four-point
correlation functions are $\lrab{\qsg\qsg\qsg\qsg}$,
$\lrab{\qsg\qsg\qs\qs}$ and $\lrab{\qs\qs\qs\qs}$.

\section{The Trispectra}

In this work, we focus on the tree-level four-point correlation
functions from direct four-point interactions. In the cosmological
context, correlation functions are conveniently calculated by using
the so-called ``in-in" formalism (see Appendix \ref{appsec_inin} for
a brief review). For our purpose, the tree-level four-point
functions are evaluated in the following form
\eq{{\label{inin_general}}
    \lrab{ \mathcal{O}(\eta_{\ast}) } = -2\,\Re \lrsb{ i \, \int^{\eta_{\ast}}_{{ -\infty} } d\eta'\,
         \soev{0}{ \mathcal{O}(\eta_{\ast})\, H_4(\eta')
         }{0}
            } \,,
         }
where $\mathcal{O}$ denotes product of four fields, e.g.
$\qsg\qsg\qsg\qsg$ etc., and $H_4$ is given in (\ref{H4_struc}).
Although we do not write down them explicitly, we should keep in
mind that, all quantities in (\ref{inin_general}) are
``interaction-picture" quantities, and thus $|0\ra$ is the free
vacuum.

\subsection{Four-point functions of the inflaton fields}

Since there is no tree-level two-point cross correlation between
$\qsg$ and $\qs$, i.e. $\lrab{ \qsg \qs }\equiv 0$, the tree-level
four-point functions are directly related to the corresponding
four-point interaction vertices. The various coefficients in
(\ref{H4_sigma})-(\ref{H4_cross}) act as ``effective couplings" of
four-point interactions. From Appendix \ref{appsec_H4_coeff}, they
are combinations of $H$, $\epsilon$, $\ca$, $\ce$, $\lambda$ and
$\Pi$, thus in this work, we treat them as approximately constant.

\subsubsection{$ \lrab{ \qsg\qsg\qsg\qsg} $}

\begin{figure}[h]
\centering
    \begin{minipage}{0.8\textwidth}
        \centering
            \begin{minipage}{0.28\textwidth}
                \centering
                \includegraphics[width=2.5cm]{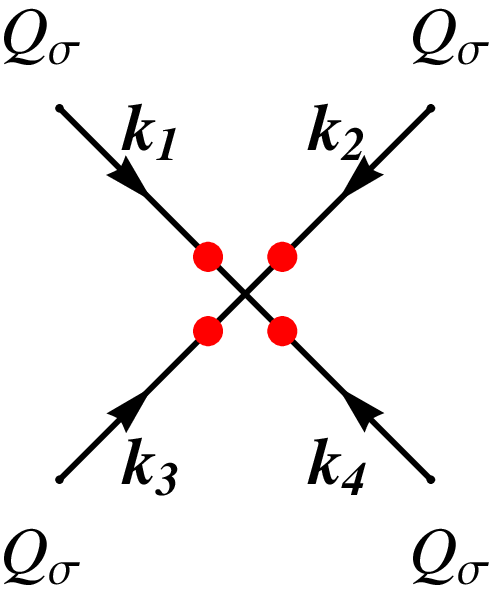}
            \end{minipage}
            \begin{minipage}{0.28\textwidth}
                \centering
                \includegraphics[width=2.5cm]{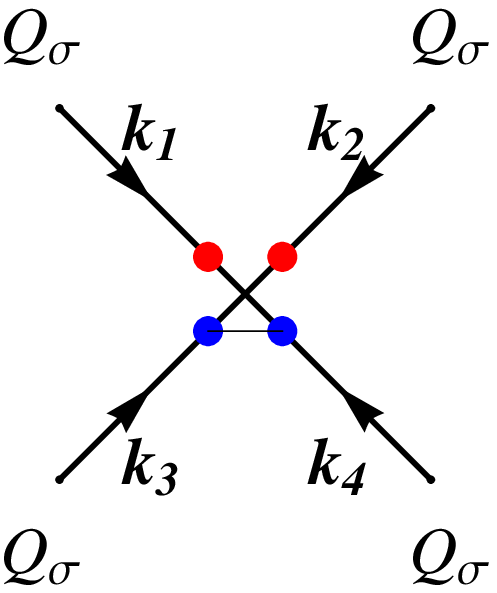}
            \end{minipage}
            \begin{minipage}{0.28\textwidth}
                \centering
                \includegraphics[width=2.5cm]{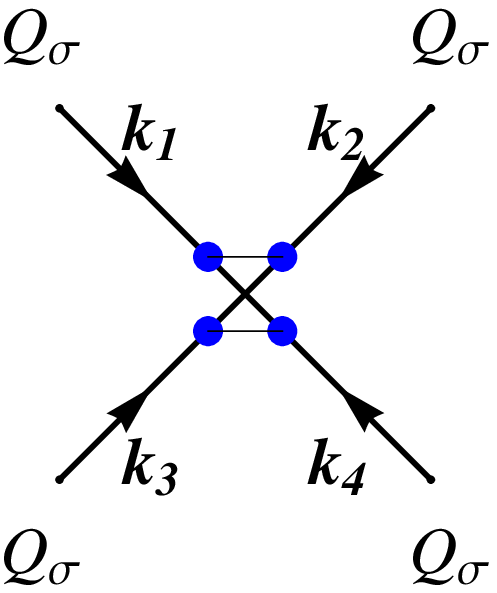}
            \end{minipage}
            \caption{Diagrammatic representations of the four-point adiabatic mode self-interaction vertices. A ``red dot" denotes derivative with respect to comoving time $\eta$.
            A blue dot denotes derivative with respect to spatial coordinates, or in Fourier space the spatial momentum. A line between two blue dots denotes ``scalar-product".}
        \label{fig_4adibatic}
        \end{minipage}
    \end{figure}

There are three types of four-point adiabatic mode self-interaction
vertices, as shown in fig.\ref{fig_4adibatic}.
And it is easy to read the corresponding contributions according to
the Feynman-diagram-like representations in fig.\ref{fig_4adibatic}:
\ea{
    \lrab{\qsg\qsg\qsg\qsg}_1 &\equiv (2\pi)^3 \delta^3(\vk_{1234})
    \lrp{-2\, \Re} \lrsb{ 24\, \Gamma_{\sigma}\, i \int_{-\infty}^{\eta_{\ast}} d\eta'\, \od{}{\eta} G_{k_1}(\eta_{\ast},\eta) \od{}{\eta} G_{k_2}(\eta_{\ast},\eta) \od{}{\eta} G_{k_3}(\eta_{\ast},\eta) \od{}{\eta} G_{k_4}(\eta_{\ast},\eta)
    } \\
    &=  (2\pi)^3 \delta^3(\vk_{1234}) \Gamma _{\sigma } \lrp{ -\frac{72 H^8  }{c_a \prod_i k_i^3 }
    } \frac{\prod_i k_i^2  }{ K^5  } \,,
} where we define $K \equiv k_1 + k_2 + k_3 +k_4$ for short.
 \ea{
    &\quad\; \lrab{\qsg\qsg\qsg\qsg}_2\\
     &\equiv (2\pi)^3 \delta^3(\vk_{1234})
    \lrp{-2\, \Re} \lrsb{ 4\, \Theta_{\sigma}\, \lrp{-\vk_3\cdot \vk_4} i \int_{-\infty}^{\eta_{\ast}} d\eta'\, \od{}{\eta} G_{k_1}(\eta_{\ast},\eta) \od{}{\eta} G_{k_2}(\eta_{\ast},\eta)  G_{k_3}(\eta_{\ast},\eta) G_{k_4}(\eta_{\ast},\eta)
    +\textrm{5perms }} \\
    &=  (2\pi)^3 \delta^3(\vk_{1234}) \Theta_{\sigma} \lrp{ \frac{H^8 }{ c_a^3 \prod_i k_i^3 }  } \lrsb{ \lrp{-\vk_3\cdot \vk_4} \frac{ k_1^2 k_2^2}{K^3 } \lrp{ 1+12 \frac{k_3 k_4}{K^2} + 3  \frac{k_3+k_4}{K} } +\textrm{5perms}
    } \,,
} where ``5 perms" denotes other 5 possibilities of choosing two out
of the four momenta $\vk_1\cdots \vk_4$ as scalar-products.
 \ea{
    &\quad\; \lrab{\qsg\qsg\qsg\qsg}_3 \\
    &\equiv (2\pi)^3 \delta^3(\vk_{1234})
    \lrp{-2\, \Re} \lrsb{ 8 \lrp{\vk_1 \cdot \vk_2 }\lrp{\vk_3 \cdot \vk_4}\, \Omega_{\sigma}\, i \int_{-\infty}^{\eta_{\ast}} d\eta'\,  G_{k_1}(\eta_{\ast},\eta)  G_{k_2}(\eta_{\ast},\eta)  G_{k_3}(\eta_{\ast},\eta)  G_{k_4}(\eta_{\ast},\eta)
    +2\textrm{perms}} \\
    &=  (2\pi)^3 \delta^3(\vk_{1234}) \Omega_{\sigma} \lrp{ \frac{ -2H^8 }{ c_a^5 \prod_i k_i^3 } } \lrsb{  \frac{ \lrp{\vk_1 \cdot \vk_2 }\lrp{\vk_3 \cdot \vk_4} }{K} \lrp{ 1 + 12 \frac{\prod_i k_i}{K^4} +  \frac{ \prod_{i<j}k_ik_j}{K^2} + 3 \frac{ \prod_i k_i}{K^3} \lrp{ \sum_i\frac{1}{k_i} }  }  +2\textrm{perms}
    }\,,
} where the other 2 permutations are combinations correspond to
scalar-products $\lrp{\vk_1 \cdot \vk_3 }\lrp{\vk_2 \cdot \vk_4}$
and $\lrp{\vk_1 \cdot \vk_4 }\lrp{\vk_2 \cdot \vk_3}$.

 It will prove convenient to define three fundamental shape
functions{\footnote{These three shape functions have been found by
several authors in investigation of trispectrum in general
single-field models \cite{Huang:2006eh,Arroja:2008yy,Chen:2009bc}.
Here in this work, for later convenience, we do not include the
permutations in the definition of $A_i$'s themselves directly, in
order to generalize them to the case of multi-field case with $\ca
\neq \ce$. The reason is that in single-field case, the shape
functions have apparent permutation symmetries, which are broken in
multi-field models.}}:
    \ea{{\label{basic_shapes}}
        A_1(k_1,k_2;k_3,k_4) &\equiv  \frac{\prod_i k_i^2  }{ K^5  } \,,\\
        A_2(k_1,k_2;\vk_3,\vk_4) &\equiv  \lrp{-\vk_3\cdot \vk_4} \frac{ k_1^2 k_2^2}{K^3 } \lrp{ 1+12 \frac{k_3 k_4}{K^2} + 3  \frac{k_3+k_4}{K} }   \,,\\
        A_3(\vk_1,\vk_2;\vk_3,\vk_4) &\equiv  \frac{ \lrp{\vk_1 \cdot \vk_2 }\lrp{\vk_3 \cdot \vk_4} }{K} \lrsb{ 1 + 12 \frac{\prod_i^4 k_i }{K^4} +  \frac{ \prod_{i<j}k_ik_j}{K^2} + 3 \frac{ \prod_i^4 k_i \lrp{ \sum_i\frac{1}{k_i} } }{K^3}   }   \,.
    }
It is useful to note two properties of these shape functions:
    \itm{
    \item permutation symmetries:\\
$A_1(k_1,k_2;k_3,k_4)$ is completely symmetric with respect to the
four momentum $k_1$, $k_2$, $k_3$ and $k_4$.
$A_2(k_1,k_2;\vk_3,\vk_4)$ is symmetric under permutations
$k_1\leftrightarrow k_2$ and $\vk_3 \leftrightarrow \vk_4 $.
Similarly, $A_3(\vk_1,\vk_2;\vk_3,\vk_4)$ is symmetric under
permutations of $\vk_1\leftrightarrow\vk_2$,
$\vk_3\leftrightarrow\vk_4 $ and $\lrp{\vk_1,\vk_2} \leftrightarrow
\lrp{\vk_3,\vk_4}$.
    \item scaling properties:\\
    \ea{
        A_1\lrp{\lambda k_1, \lambda k_2; \lambda k_3, \lambda
        k_4} = \lambda^3 {A}_1 \lrp{ k_1, k_2,
        k_3,k_4} \,,\\
        {A}_2\lrp{\lambda k_1, \lambda k_2; \lambda \vk_3, \lambda
        \vk_4} = \lambda^3 {A}_2 \lrp{ k_1, k_2,
        \vk_3,\vk_4} \,,\\
        {A}_3\lrp{\lambda \vk_1, \lambda \vk_2; \lambda \vk_3, \lambda
        \vk_4} = \lambda^3 {A}_3 \lrp{ \vk_1, \vk_2,
        \vk_3,\vk_4} \,.
    }
    In other words, $A_i$'s have momentum dimension as $k^3$:
    $[A_i]=[k^3]$.
 }

In terms of these three shape functions, we can write{\footnote{In
the last line of (\ref{4a_1}), we abstract a factor
$\mathcal{P}_{\sigma}^4$ rather than $\mathcal{P}^3$ in previous
works, in order to make the expression more symmetric. The reason
will become clearer in the following calculations for
$\lrab{\qsg\qsg\qs\qs}$.}}
    \ea{{\label{4a_1}}
       \lrab{\qsg\qsg\qsg\qsg}_1 &= (2\pi)^3 \delta^3(\vk_{1234}) \Gamma _{\sigma } \lrp{ -\frac{72 H^8  }{c_a \prod_i k_i^3 }
    } {A}_1(k_1,k_2;k_3,k_4) \\
    &\equiv  (2\pi)^{11} \delta^3(\vk_{1234}) \frac{\mathcal{P}_{\sigma\ast}^4}{\prod_i k_i^3} \lrp{-72\Gamma_{\sigma}} {A}_1\lrp{\ca k_1,
\ca k_2; \ca k_3, \ca k_4} \,,
    }
where we have used the scaling property ${A}_1\lrp{\ca k_1, \ca k_2,
\ca k_3, \ca k_4} = \ca^3 {A}_1\lrp{k_1,k_2,k_3,k_4}$. It will prove
convenient to use the ``sound speed dependent" (in the later we
denote ``$c_s$-dependent" for short) shape functions ${A}_1\lrp{\ca
k_1, \ca k_2, \ca k_3, \ca k_4}$, especially when the adiabatic mode
and entropy mode have different sound speeds: $\ca \neq \ce$. Using
the ``$c_s$-dependent shape functions" not only makes the
expressions simpler and more symmetric but also make the physical
picture clear.

Similarly, we have
    \ea{{\label{4a_23}}
        \lrab{\qsg\qsg\qsg\qsg}_2 &= (2\pi)^{11} \delta^3(\vk_{1234}) \frac{\mathcal{P}_{\sigma\ast}^4}{\prod_i
        k_i^3}\, \Theta_{\sigma} \frac{1}{\ca^2} \lrsb{ {A}_2(\ca k_1, \ca k_2; \ca \vk_3, \ca \vk_4) +\textrm{5 perms} } \\
        \lrab{\qsg\qsg\qsg\qsg}_3 &= (2\pi)^{11} \delta^3(\vk_{1234}) \frac{\mathcal{P}_{\sigma\ast}^4}{\prod_i
        k_i^3}\, \lrp{-2\Omega_{\sigma}} \frac{1}{\ca^4} \lrsb{ {A}_3(\ca \vk_1, \ca \vk_2; \ca \vk_3, \ca \vk_4) +\textrm{2 perms} } \,.
    }
The whole contribution from $\lrab{\qsg\qsg\qsg\qsg}$ is given by
    \ea{
        \lrab{\qsg\qsg\qsg\qsg} = \sum_i^3 \lrab{\qsg\qsg\qsg\qsg}_i
        \,.
    }

\subsubsection{$ \lrab{ \qsg\qsg\qs\qs} $}

\begin{figure}[h]
\centering
    \begin{minipage}{0.8\textwidth}
        \centering
            \begin{minipage}{0.28\textwidth}
                \centering
                \includegraphics[width=2.5cm]{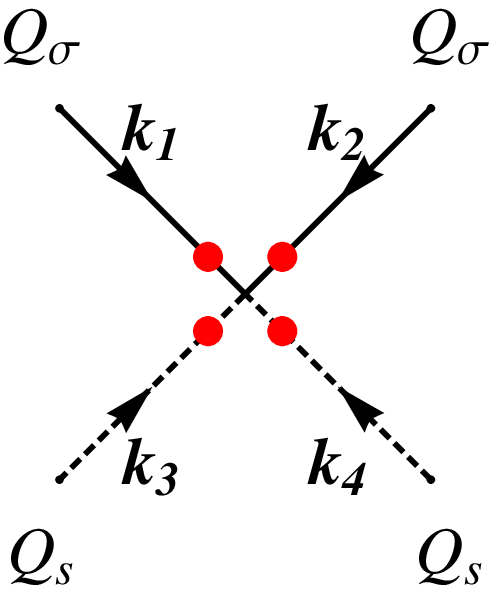}
            \end{minipage}
            \begin{minipage}{0.28\textwidth}
                \centering
                \includegraphics[width=2.5cm]{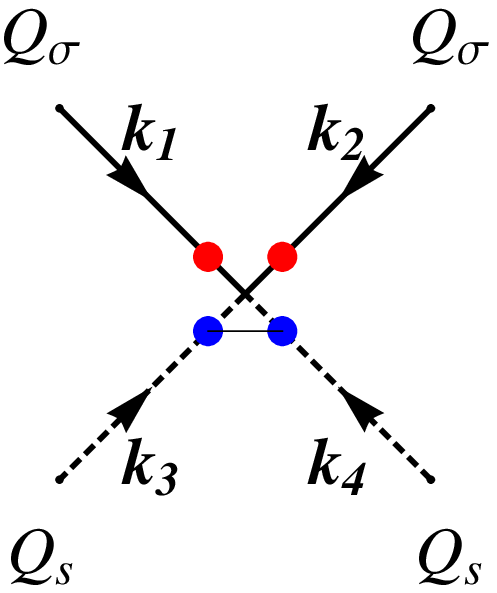}
            \end{minipage}
            \begin{minipage}{0.28\textwidth}
                \centering
                \includegraphics[width=2.5cm]{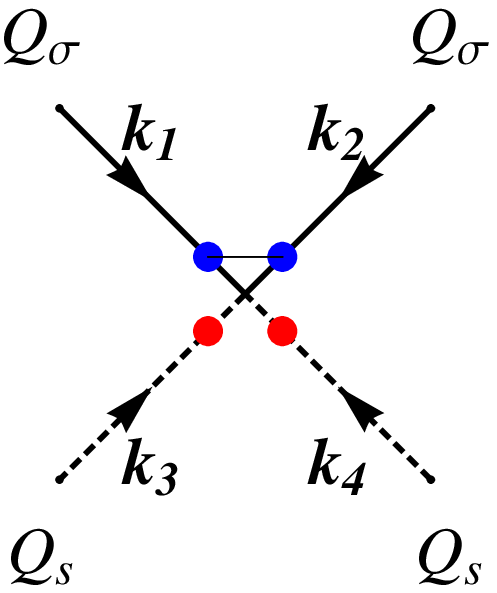}
            \end{minipage}
            \begin{minipage}{0.28\textwidth}
                \centering
                \includegraphics[width=2.5cm]{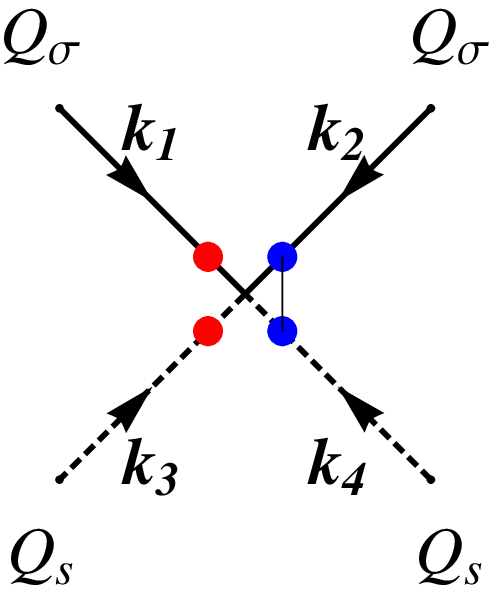}
            \end{minipage}
            \begin{minipage}{0.28\textwidth}
                \centering
                \includegraphics[width=2.5cm]{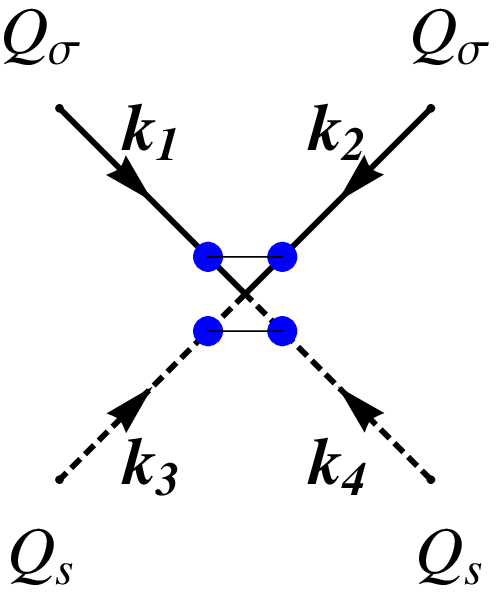}
            \end{minipage}
            \begin{minipage}{0.28\textwidth}
                \centering
                \includegraphics[width=2.5cm]{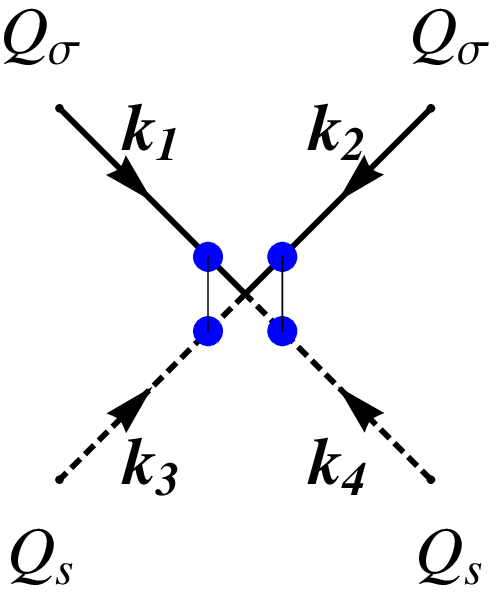}
            \end{minipage}
            \caption{Diagrammatic representations of the four-point adiabatic/entropy modes cross-interaction vertices. A ``red dot" denotes derivative wrt comoving time $\eta$.
            A blue dot donotes derivative wrt spatial coordinates, or in Fourier space the spatial momentum. A line between two blue dots denotes ``scalar-product".}
        \label{fig_4cross}
        \end{minipage}
    \end{figure}

There are six types of adiabatic/entropy four-point
cross-interaction
 vertices, involving two adiabatic modes and two entropy
modes, as shown in fig.\ref{fig_4cross}.

\ea{{\label{2a2s_1}}
    \lrab{ \qsg \qsg \qs\qs}_1 &\equiv (2\pi)^3 \delta^3(\vk_{1234}) \lrp{ -2\Re} \lrsb{ 4 \Gamma_c\, i \int_{-\infty}^{\eta_{\ast}} d\eta\, \od{}{\eta} G_{k_1}(\eta_{\ast},\eta) \od{}{\eta} G_{k_2}(\eta_{\ast},\eta) \od{}{\eta} F_{k_3}(\eta_{\ast},\eta) \od{}{\eta} F_{k_4}(\eta_{\ast},\eta)
    } \\
    &= (2\pi)^3 \delta^3(\vk_{1234}) \Gamma_c  \lrp{ \frac{- 12  H^8 }{\prod_i k_i^3 } }  \frac{ c_a^2 c_e^2 \prod_i^4 k_i^2}{ \tilde{K}^5
    } \,.
} where $\tilde{K} \equiv \ca \lrp{ k_1 +k_2 } +\ce \lrp{ k_3 +
k_4}$. In terms of the three fundamental shape functions
(\ref{basic_shapes}), (\ref{2a2s_1}) can be written in a rather
compact and convenient form:
    \eq{
        \lrab{ \qsg \qsg \qs\qs}_1 = (2\pi)^{11}
        \delta^3(\vk_{1234}) \frac{\mathcal{P}_{\sigma\ast}^2 \mathcal{P}_{s\ast}^2 }{\prod_i k_i^3
        } \lrp{-12 \Gamma_c} {A}_1 \lrp{ \ca k_1, \ca k_2; \ce k_3, \ce k_4
        } \,,
    }
where $\mathcal{P}_{\sigma}$ and $\mathcal{P}_s$ are the
dimensionless power spectrum defined in (\ref{dimless_spec}).
Similarly, we have (readers who are interested in their explicit
expressions may refer to Appendix \ref{appsec_2a2s})
    \ea{
        \lrab{ \qsg \qsg \qs\qs}_2 &= (2\pi)^{11}
        \delta^3(\vk_{1234}) \frac{\mathcal{P}_{\sigma\ast}^2 \mathcal{P}_{s\ast}^2 }{\prod_i k_i^3
        }\, \frac{\Theta_{\sigma s}}{\ce^2} {A}_2\lrp{ \ca k_1, \ca k_2; \ce \vk_3, \ce \vk_4 } \,,\\
        \lrab{ \qsg \qsg \qs\qs}_3 &= (2\pi)^{11}
        \delta^3(\vk_{1234}) \frac{\mathcal{P}_{\sigma\ast}^2 \mathcal{P}_{s\ast}^2 }{\prod_i k_i^3
        }\,  \frac{\Theta_{s\sigma}}{\ca^2} {A}_2\lrp{ \ce k_3, \ce k_4; \ca \vk_1, \ca \vk_2 } \,,\\
        \lrab{ \qsg \qsg \qs\qs}_4 &= (2\pi)^{11}
        \delta^3(\vk_{1234}) \frac{\mathcal{P}_{\sigma\ast}^2 \mathcal{P}_{s\ast}^2 }{\prod_i k_i^3
        }\,  \frac{\Theta_{c}}{4 \, \ca\ce} \left[ {A}_2 \lrp{ \ca k_1, \ce k_3; \ca \vk_2, \ce \vk_4} + {A}_2 \lrp{ \ca k_1, \ce k_4; \ca \vk_2, \ce \vk_3} \right. \\
        &\qquad\qquad\qquad\qquad\qquad\qquad\qquad \left. + {A}_2 \lrp{ \ca k_2, \ce k_3; \ca \vk_1, \ce \vk_4} + {A}_2 \lrp{ \ca k_2, \ce k_4; \ca \vk_1, \ce \vk_3} \right] \,,\\
        \lrab{ \qsg \qsg \qs\qs}_5 &= (2\pi)^{11}
        \delta^3(\vk_{1234}) \frac{\mathcal{P}_{\sigma\ast}^2 \mathcal{P}_{s\ast}^2 }{\prod_i k_i^3
        } \, \frac{\lrp{-\Omega_{\sigma s}}}{\ca^2\ce^2}\, {A}_3 \lrp{ \ca\vk_1, \ca\vk_2; \ce\vk_3,\ce\vk_4 }  \,,\\
        \lrab{ \qsg \qsg \qs\qs}_6 &= (2\pi)^{11}
        \delta^3(\vk_{1234}) \frac{\mathcal{P}_{\sigma\ast}^2 \mathcal{P}_{s\ast}^2 }{\prod_i k_i^3
        }\,  \frac{\lrp{-\Omega_c}}{2\ca^2\ce^2}\lrsb{ {A}_3\lrp{ \ca\vk_1, \ce\vk_3; \ca\vk_2, \ce\vk_4 } + {A}_3\lrp{ \ca\vk_1, \ce\vk_4; \ca\vk_2, \ce\vk_3 } } \,.
    }

For clarity and later convenience, we group these six contributions
into three types, arising from ${A}_1$, ${A}_2$ and ${A}_3$
respectively. First we have,
    \eq{{\label{2a2a_A1}}
        \lrab{\qsg \qsg \qs\qs}_{{A}_1} \equiv \lrab{\qsg \qsg
        \qs\qs}_1 = (2\pi)^{11}
        \delta^3(\vk_{1234}) \frac{\mathcal{P}_{\sigma\ast}^2 \mathcal{P}_{s\ast}^2 }{\prod_i k_i^3
        } \lrp{-12 \Gamma_c} {A}_1 \lrp{ \ca k_1, \ca k_2; \ce k_3, \ce k_4
        } \,,
    }
While
    \ea{{\label{2a2s_A2}}
        \lrab{\qsg \qsg \qs\qs}_{{A}_2} &\equiv  \lrab{\qsg \qsg
        \qs\qs}_2 + \lrab{\qsg \qsg \qs\qs}_3 + \lrab{\qsg \qsg
        \qs\qs}_4 \\
        &= (2\pi)^{11}
        \delta^3(\vk_{1234}) \frac{\mathcal{P}_{\sigma\ast}^2 \mathcal{P}_{s\ast}^2 }{\prod_i k_i^3
        }\, \left\{  \frac{\Theta_{\sigma s}}{\ce^2} {A}_2\lrp{ \ca k_1, \ca k_2; \ce \vk_3, \ce \vk_4 } + \frac{\Theta_{s\sigma}}{\ca^2} {A}_2\lrp{ \ce k_3, \ce k_4; \ca \vk_1, \ca \vk_2 }
        \right. \\
        &\qquad\qquad\qquad\qquad + \frac{\Theta_c}{4\ca\ce} \left[ {A}_2 \lrp{ \ca k_1, \ce k_3; \ca \vk_2, \ce \vk_4} + {A}_2 \lrp{ \ca k_1, \ce k_4; \ca \vk_2, \ce \vk_3} \right. \\
        &\qquad\qquad\qquad\qquad\qquad\qquad \left. + {A}_2 \lrp{ \ca k_2, \ce k_3; \ca \vk_1, \ce \vk_4} + {A}_2 \lrp{ \ca k_2, \ce k_4; \ca \vk_1, \ce \vk_3}
        \right] \Big\} \,,
    }
and
    \ea{{\label{2a2s_A3}}
        \lrab{\qsg \qsg \qs\qs}_{{A}_3} &\equiv  \lrab{\qsg \qsg
        \qs\qs}_5 + \lrab{\qsg\qsg\qs\qs}_6 \\
        &= (2\pi)^{11}
        \delta^3(\vk_{1234}) \frac{\mathcal{P}_{\sigma\ast}^2 \mathcal{P}_{s\ast}^2 }{\prod_i k_i^3
        }\,  \left\{  \frac{\lrp{-\Omega_{\sigma s}}}{\ca^2\ce^2}\, {A}_3 \lrp{ \ca\vk_1, \ca\vk_2; \ce\vk_3,\ce\vk_4 }  \right. \\
        &\qquad\qquad\qquad\qquad \left. + \frac{\lrp{-\Omega_c}}{2\ca^2\ce^2}\lrsb{ \mathcal{A}_3\lrp{ \ca\vk_1, \ce\vk_3; \ca\vk_2, \ce\vk_4 } + {A}_3\lrp{ \ca\vk_1, \ce\vk_4; \ca\vk_2, \ce\vk_3 } } \right\} \,.
    }
The whole contribution from $\lrab{\qsg\qsg\qs\qs}$ is given by
 \ea{
    \lrab{ \qsg(\vk_1,\eta_{\ast}) \qsg(\vk_2,\eta_{\ast}) \qs(\vk_3,\eta_{\ast}) \qs(\vk_4,\eta_{\ast})
    } \equiv \lrab{\qsg\qsg\qs\qs}_{{A}_1} + \lrab{\qsg\qsg\qs\qs}_{{A}_2} + \lrab{\qsg\qsg\qs\qs}_{{A}_3} \,.
}

\subsubsection{$ \lrab{ \qs \qs\qs\qs} $}

There are also three types of $\qs$ self-interaction four-point
vertices, as shown in fig.\ref{fig_4entropy}. $ \lrab{ \qs
\qs\qs\qs} $ are easily obtained by simply changing the corresponding
``effective couplings" and $\mathcal{P}_{\sigma} \leftrightarrow
\mathcal{P}_s$ and  $\ca \leftrightarrow \ce$ in
(\ref{4a_1})-(\ref{4a_23}):
\begin{figure}[h]
\centering
    \begin{minipage}{0.8\textwidth}
        \centering
            \begin{minipage}{0.28\textwidth}
                \centering
                \includegraphics[width=2.5cm]{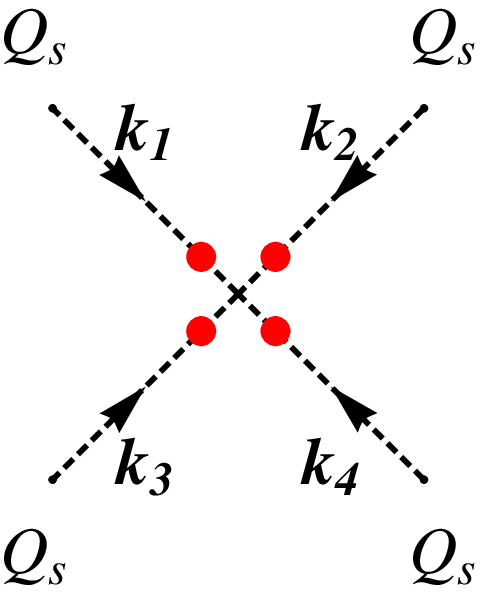}
            \end{minipage}
            \begin{minipage}{0.28\textwidth}
                \centering
                \includegraphics[width=2.5cm]{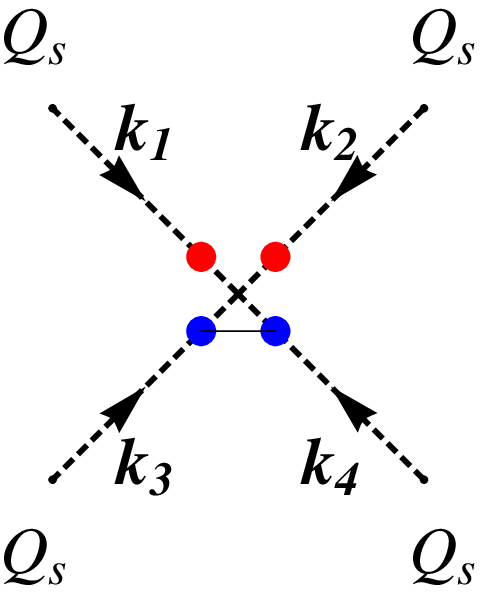}
            \end{minipage}
            \begin{minipage}{0.28\textwidth}
                \centering
                \includegraphics[width=2.5cm]{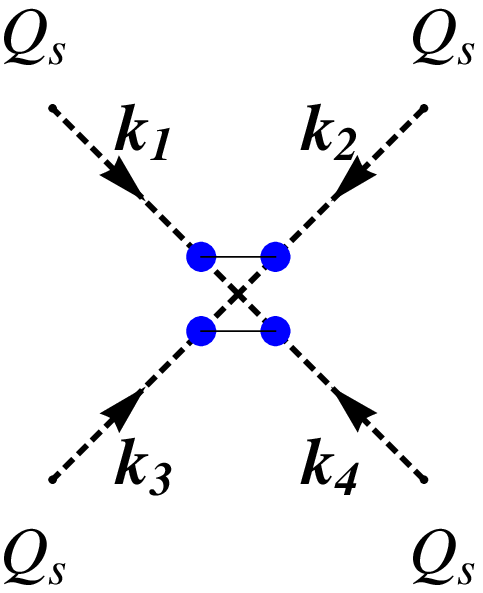}
            \end{minipage}
            \caption{Diagrammatic representations of the four-point entropy mode self-interaction vertices. A ``red dot" denotes derivative wrt comoving time $\eta$. A blut dot donotes derivative wrt spatial coordinates, or in fourier space the spatial momentum. A line between two blue dots denotes ``scalar-product".}
        \label{fig_4entropy}
        \end{minipage}
    \end{figure}
    \ea{{\label{4s_final}}
        \lrab{\qs\qs\qs\qs}_1 &=   (2\pi)^{11} \delta^3(\vk_{1234}) \frac{\mathcal{P}_{s\ast}^4}{\prod_i^4 k_i^3} \lrp{-72\Gamma_{s}} {A}_1\lrp{\ce k_1,
\ce k_2, \ce k_3, \ce k_4} \,,\\
        \lrab{\qs\qs\qs\qs}_2 &= (2\pi)^{11} \delta^3(\vk_{1234}) \frac{\mathcal{P}_{s\ast}^4}{\prod_i^4
        k_i^3}\, \Theta_{s} \frac{1}{\ce^2} \lrsb{ {A}_2(\ce k_1, \ce k_2; \ce \vk_3, \ce \vk_4) +\textrm{5 perms} } \\
        \lrab{\qs\qs\qs\qs}_3 &= (2\pi)^{11} \delta^3(\vk_{1234}) \frac{\mathcal{P}_{s\ast}^4}{\prod_i^4
        k_i^3}\, \lrp{-2\Omega_{s}} \frac{1}{\ce^4} \lrsb{ {A}_3(\ce \vk_1, \ce \vk_2; \ce \vk_3, \ce \vk_4) +\textrm{2 perms} } \,.
    }
The whole contribution from $\lrab{\qs\qs\qs\qs}$ is given by
 \ea{
    &\quad\; \lrab{ \qs(\vk_1,\eta_{\ast}) \qs(\vk_2,\eta_{\ast}) \qs(\vk_3,\eta_{\ast}) \qs(\vk_4,\eta_{\ast})
    } = \sum_i^3 \lrab{\qs\qs\qs\qs}_i \,.
}

\subsection{Four-point Function of the Curvature Perturbation $\lrab{\mathcal{R}^4}$}

As we have addressed before,  the scalar-field perturbations $\qsg$
and $\qs$ themselves are not directly observable. What we are
eventually interested in is the curvature perturbation
$\mathcal{R}$. As has been investigated in
\cite{Langlois:2008qf,Langlois:2008wt,Arroja:2008yy}, this can be
achieved by writing
    \ea{
        \mathcal{R} &\approx \mathcal{R}_{\ast} +
        T_{\mathcal{R}\mathcal{S}} \mathcal{S}_{\ast} =
        \lrp{\frac{H}{\dot{\sigma}}}_{\ast} Q_{\sigma\ast} +
        T_{\mathcal{R}\mathcal{S}}\lrp{\frac{H}{\dot{\sigma}}}_{\ast}
        Q_{s\ast} \\
        &\equiv \mathcal{N}_{\sigma} Q_{\sigma\ast} + \mathcal{N}_s
        Q_{s\ast} \,.
    }
Note that $\mathcal{N}_s \equiv T_{\mathcal{R}\mathcal{S}}
\mathcal{N}_{\sigma}$ is in general time-dependent. The four-point
correlation function for comoving curvature perturbation
$\mathcal{R}$ is given by
    \ea{{\label{4R_original}}
        &\quad\; \lrab{ \Rc(\vk_1)\Rc(\vk_3)\Rc(\vk_2)\Rc(\vk_4) } \\
        &=
        \mathcal{N}_{\sigma}^4 \lrab{ \qsg(\vk_1)\qsg(\vk_2)\qsg(\vk_3)\qsg(\vk_4)
        }_{\ast} + \mathcal{N}_{\sigma}^2 \mathcal{N}_s^2
        \lrsb{ \lrab{ \qsg(\vk_1)\qsg(\vk_2)\qs(\vk_3)\qs(\vk_4) }_{\ast} + \textrm{5 perms} }  \\
        &\qquad\qquad + \mathcal{N}_{s}^4 \lrab{ \qs(\vk_1)\qs(\vk_2)\qs(\vk_3)\qs(\vk_4)
        }_{\ast} \,,
    }
where ``5 perms" denotes other 5 possibilities of choosing two
momenta for $\qsg$ and two momenta for $\qs$ out of the four
momenta{\footnote{That is, $\lrab{
\qsg(\vk_1)\qsg(\vk_3)\qs(\vk_2)\qs(\vk_4) }$, $\lrab{
\qsg(\vk_1)\qsg(\vk_4)\qs(\vk_2)\qs(\vk_3) }$, $\lrab{
\qsg(\vk_2)\qsg(\vk_3)\qs(\vk_1)\qs(\vk_4) }$, $\lrab{
\qsg(\vk_2)\qsg(\vk_4)\qs(\vk_1)\qs(\vk_3) }$ and $\lrab{
\qsg(\vk_3)\qsg(\vk_4)\qs(\vk_1)\qs(\vk_2) }$.}}. In deriving
(\ref{4R_original}), we have used the assumption that there is no
cross-correlation between adiabatic and entropy modes around
horizon-crossing, i.e. $\lrab{\qsg\qs}_{\ast}\equiv 0$. This is
indeed the case if we use the second-order perturbation action
(\ref{S2_Q}) as our starting point of canonical quantization.

It is also convenient to group the whole contributions to
$\lrab{\mathcal{R}^4}$ (\ref{4R_original}) into three types, which
correspond to three different physical origins or more precisely
three different types of four-point interaction vertices. From
(\ref{4a_1}), (\ref{2a2a_A1}) and (\ref{4s_final}), we have
\ea{{\label{4R_A1}}
    \lrab{\Rc^4}_{1} &\equiv (2\pi)^{11} \delta^3(\vk_{1234})
    \frac{1}{\prod_i k_i^3} \left\{ \mathcal{N}_{\sigma}^4 \mathcal{P}_{\sigma\ast}^4 \lrp{-12\Gamma_{\sigma}} {A}_1\lrp{\ca k_1,
\ca k_2; \ca k_3, \ca k_4} \right. \\
    &\qquad\qquad + \mathcal{N}_{\sigma}^2\mathcal{N}_{s}^2 \mathcal{P}_{\sigma\ast}^2 \mathcal{P}_{s\ast}^2 \, \lrp{-12 \Gamma_c} {A}_1 \lrp{ \ca k_1, \ca k_2; \ce k_3, \ce k_4
        } \\
    &\qquad\qquad \left.+ \mathcal{N}_{s}^4 \mathcal{P}_{s\ast}^4\, \lrp{-12\Gamma_{s}} {A}_1\lrp{\ce k_1,
\ce k_2, \ce k_3, \ce k_4} +\textrm{5 perms}  \right\} \\
    &= (2\pi)^{11} \delta^3(\vk_{1234})
    \frac{-12 \mathcal{N}_{\sigma}^4 \mathcal{P}_{\sigma\ast}^4 }{\prod_i k_i^3} \bigg\{ \Gamma_{\sigma}\, {A}_1\lrp{\ca k_1,
\ca k_2; \ca k_3, \ca k_4}  \\
    &\qquad\qquad + \lrp{ T_{\mathcal{R}\mathcal{S}} \frac{\ca}{\ce}}^2 \, \Gamma_c\, {A}_1 \lrp{ \ca k_1, \ca k_2; \ce k_3, \ce k_4
        } \\
    &\qquad\qquad \left.+ \lrp{ T_{\mathcal{R}\mathcal{S}}  \frac{\ca}{\ce}}^4\, \Gamma_{s}\, {A}_1\lrp{\ce k_1,
\ce k_2, \ce k_3, \ce k_4} +\textrm{5 perms}  \right\} \,,
}
    where we have used $\mathcal{N}_s \equiv T_{\mathcal{R}\mathcal{S}}
\mathcal{N}_{\sigma}$ and the fact that $\mathcal{P}_s /
\mathcal{P}_{\sigma} =
    \ca/\ce$. ``5 perms" denotes other 5 possibilities of choosing two out from the four momenta $k_1,\cdots, k_4$ as the first two arguments of ${A}_1$ and the other two momenta as the last two arguments of ${A}_1$(obviously the differences only arise in ${A}_1 \lrp{ \ca k_1, \ca k_2; \ce k_3, \ce k_4}$). Similarly, we have
    \ea{{\label{4R_A2}}
        \lrab{\Rc^4}_{2} &=  (2\pi)^{11} \delta^3(\vk_{1234})
    \frac{\mathcal{N}_{\sigma}^4 \mathcal{P}_{\sigma\ast}^4  }{\prod_i k_i^3} \left\{  \frac{ \Theta_{\sigma}}{\ca^2}  {A}_2(\ca k_1, \ca k_2; \ca \vk_3, \ca \vk_4)
    \right.\\
    &\qquad\qquad + \lrp{ T_{\mathcal{R}\mathcal{S}} \frac{\ca}{\ce}}^2  \left[ \frac{\Theta_{\sigma s}}{\ce^2} {A}_2\lrp{ \ca k_1, \ca k_2; \ce \vk_3, \ce \vk_4 } + \frac{\Theta_{s\sigma}}{\ca^2} {A}_2\lrp{ \ce k_3, \ce k_4; \ca \vk_1, \ca \vk_2 }
        \right. \\
        &\qquad\qquad\qquad\qquad + \frac{\Theta_c}{4\ca\ce} \left[ {A}_2 \lrp{ \ca k_1, \ce k_3; \ca \vk_2, \ce \vk_4} + {A}_2 \lrp{ \ca k_1, \ce k_4; \ca \vk_2, \ce \vk_3} \right. \\
        &\qquad\qquad\qquad\qquad\qquad\qquad \left. + {A}_2 \lrp{ \ca k_2, \ce k_3; \ca \vk_1, \ce \vk_4} + {A}_2 \lrp{ \ca k_2, \ce k_4; \ca \vk_1, \ce \vk_3}
        \right] \bigg]\\
        &\qquad\qquad + \lrp{ T_{\mathcal{R}\mathcal{S}} \frac{\ca}{\ce}}^4  \frac{\Theta_{s}}{\ce^2}  {A}_2(\ce k_1, \ce k_2; \ce \vk_3, \ce
        \vk_4) +\textrm{5 perms} \bigg\}\,,
    }
and
    \ea{{\label{4R_A3}}
        \lrab{\mathcal{R}^4}_{3} &= (2\pi)^{11} \delta^3(\vk_{1234})
    \frac{\mathcal{N}_{\sigma}^4 \mathcal{P}_{\sigma\ast}^4 }{\prod_i k_i^3} \left\{  \frac{\lrp{-\Omega_{\sigma}}}{\ca^4}  {A}_3(\ca \vk_1, \ca \vk_2; \ca \vk_3, \ca \vk_4)
    \right.\\
    &\qquad\qquad\qquad\qquad  + \lrp{ T_{\mathcal{R}\mathcal{S}} \frac{\ca}{\ce}}^2 \left[  \frac{\lrp{-\Omega_{\sigma s}}}{\ca^2\ce^2}\, {A}_3 \lrp{ \ca\vk_1, \ca\vk_2; \ce\vk_3,\ce\vk_4 }  \right. \\
        &\qquad\qquad\qquad\qquad\qquad\qquad \left. + \frac{\lrp{-\Omega_c}}{2\ca^2\ce^2}\lrsb{ {A}_3\lrp{ \ca\vk_1, \ce\vk_3; \ca\vk_2, \ce\vk_4 } + {A}_3\lrp{ \ca\vk_1, \ce\vk_4; \ca\vk_2, \ce\vk_3 } }
        \right] \\
        &\qquad\qquad\qquad\qquad \left. + \lrp{ T_{\mathcal{R}\mathcal{S}} \frac{\ca}{\ce}}^4  \frac{\lrp{-\Omega_{s}}}{\ce^4} {A}_3(\ce \vk_1, \ce \vk_2; \ce \vk_3, \ce
        \vk_4) + \textrm{5 perms} \right\} \,.
    }

It is convenient to define three new shape functions, which can be
viewed as the generalizations of the three fundamental shapes
(\ref{basic_shapes}).
    \ea{{\label{new_A1}}
        \mathcal{A}_1\lrp{k_1,k_2,k_3,k_3} &\equiv   {A}_1\lrp{\ca k_1,
\ca k_2; \ca k_3, \ca k_4}   + \lrp{ T_{\mathcal{R}\mathcal{S}}
\frac{\ca}{\ce}}^2 \, \frac{\Gamma_c}{\Gamma_{\sigma}}\, {A}_1 \lrp{
\ca k_1, \ca k_2; \ce k_3, \ce k_4
        } \\
    &\qquad\qquad + \lrp{ T_{\mathcal{R}\mathcal{S}}  \frac{\ca}{\ce}}^4\, \frac{\Gamma_{s}}{\Gamma_{\sigma}}\, {A}_1\lrp{\ce k_1,
\ce k_2, \ce k_3, \ce k_4} +\textrm{5 perms} \,,
    }
    \ea{{\label{new_A2}}
        \mathcal{A}_2(\vk_1,\vk_2;\vk_3,\vk_4) &\equiv   \frac{ 1}{\ca^2}  {A}_2(\ca k_1, \ca k_2; \ca \vk_3, \ca \vk_4) \\
    &\qquad\qquad + \lrp{ T_{\mathcal{R}\mathcal{S}} \frac{\ca}{\ce}}^2  \frac{1}{\Theta_{\sigma}}\left[ \frac{\Theta_{\sigma s}}{\ce^2} {A}_2\lrp{ \ca k_1, \ca k_2; \ce \vk_3, \ce \vk_4 } + \frac{\Theta_{s\sigma}}{\ca^2} {A}_2\lrp{ \ce k_3, \ce k_4; \ca \vk_1, \ca \vk_2 }
        \right. \\
        &\qquad\qquad\qquad\qquad + \frac{\Theta_c}{4\ca\ce} \left[ {A}_2 \lrp{ \ca k_1, \ce k_3; \ca \vk_2, \ce \vk_4} + {A}_2 \lrp{ \ca k_1, \ce k_4; \ca \vk_2, \ce \vk_3} \right. \\
        &\qquad\qquad\qquad\qquad\qquad\qquad \left. + {A}_2 \lrp{ \ca k_2, \ce k_3; \ca \vk_1, \ce \vk_4} + {A}_2 \lrp{ \ca k_2, \ce k_4; \ca \vk_1, \ce \vk_3}
        \right] \bigg]\\
        &\qquad\qquad + \lrp{ T_{\mathcal{R}\mathcal{S}} \frac{\ca}{\ce}}^4  \frac{\Theta_{s}}{\Theta_{\sigma} \ce^2}  {A}_2(\ce k_1, \ce k_2; \ce \vk_3, \ce
        \vk_4) +\textrm{5 perms}\,,
    }
and
    \ea{{\label{new_A3}}
        \mathcal{A}_3(\vk_1,\vk_2;\vk_3,\vk_4) &\equiv   \frac{1}{\ca^4}  {A}_3(\ca \vk_1, \ca \vk_2; \ca \vk_3, \ca \vk_4) \\
    &\qquad\qquad  + \lrp{ T_{\mathcal{R}\mathcal{S}} \frac{\ca}{\ce}}^2 \frac{1}{\Omega_{\sigma}}\left[  \frac{ \Omega_{\sigma s}}{\ca^2\ce^2}\, {A}_3 \lrp{ \ca\vk_1, \ca\vk_2; \ce\vk_3,\ce\vk_4 }  \right. \\
        &\qquad\qquad\qquad\qquad \left. + \frac{\Omega_c}{2\ca^2\ce^2}\lrsb{ {A}_3\lrp{ \ca\vk_1, \ce\vk_3; \ca\vk_2, \ce\vk_4 } + {A}_3\lrp{ \ca\vk_1, \ce\vk_4; \ca\vk_2, \ce\vk_3 } }
        \right] \\
        &\qquad\qquad  + \lrp{ T_{\mathcal{R}\mathcal{S}} \frac{\ca}{\ce}}^4  \frac{\Omega_{s}}{\Omega_{\sigma} \ce^4} {A}_3(\ce \vk_1, \ce \vk_2; \ce \vk_3, \ce
        \vk_4) +\textrm{5 perms}\,.
    }
Note that $\mathcal{A}_i$ also depend on parameters $\ca$, $\ce$,
$\lambda$, $\Pi$ and $T_{\Rc\Sc}$. In multi-field models, these
parameters enter the definition of the shapes. This is essentially
different from that in single-field models, we can always abstract
shape functions, which are functions of momenta only. In terms of
these three ``generalized shape functions"
(\ref{new_A1})-(\ref{new_A3}), the four-point correlation function
for comoving curvature perturbation $\mathcal{R}$ can be recast into
a rather convenient form:
    \ea{{\label{4R_final}}
        \lrab{ \mathcal{R}(\vk_1)\mathcal{R}(\vk_2)\mathcal{R}(\vk_3)\mathcal{R}(\vk_4)
        } = (2\pi)^{11} \delta^3(\vk_{1234}) \frac{\mathcal{N}_{\sigma}^4\mathcal{P}_{\sigma\ast}^4}{\prod_i k_i^3} \lrsb{ -12\Gamma_{\sigma} \mathcal{A}_1 + \Theta_{\sigma}\mathcal{A}_2 - \Omega_{\sigma}\mathcal{A}_3
        }\,.
    }

\subsubsection{Trispectrum of $\mathcal{R}$}{\label{sec_tri}}

In practice, it is convenient to define a so-called trispectrum for
$\mathcal{R}$:
    \eq{{\label{trispectrum_def}}
        \lrab{ \mathcal{R}(\vk_1)\mathcal{R}(\vk_2)\mathcal{R}(\vk_3)\mathcal{R}(\vk_4)
        } \equiv (2\pi)^3 \delta^3(\vk_{1234})
        T_{\Rc}(\vk_1,\vk_2,\vk_3,\vk_4) \,.
    }
From (\ref{4R_final}) we have
    \ea{{\label{trispectrum_R}}
        T_{\Rc}(\vk_1,\vk_2,\vk_3,\vk_4) &= \frac{H^8}{4\epsilon^2
        \ca^4} \frac{1}{\prod_i k_i^3} \lrp{ -12\Gamma_{\sigma} \mathcal{A}_1 + \Theta_{\sigma}\mathcal{A}_2 - \Omega_{\sigma}\mathcal{A}_3
        } \\
        &= (2\pi)^6
        \mathcal{P}_{\Rc\ast}^3 \frac{2H^2\epsilon}{\ca \prod_i k_i^3} \lrp{ -12\Gamma_{\sigma} \mathcal{A}_1 + \Theta_{\sigma}\mathcal{A}_2 - \Omega_{\sigma}\mathcal{A}_3
        } \,,
    }
where we have used $\mathcal{N}_{\sigma}^2 = 1/2\epsilon$ and
$\mathcal{P}_{\sigma\ast} = \frac{1}{\ca}\lrp{\frac{H}{2\pi}}^2$. At
this point, it is useful to note the dimensions of various
quantities:
    \itm{
        \item $[\Gamma_{\sigma}]=[\Theta_{\sigma}]=[\Omega_{\sigma}] =
        [1/H^2]$,
        \item
        $[\mathcal{A}_1]=[\mathcal{A}_2]=[\mathcal{A}_3]=[k^3]$,
    \item thus, the trispectrum has dimension as $[T_{\Rc}] =[H^6/k^9]$. (Recall that the power spectrum has dimension $[P_{\Rc}]=[H^2/k^3]$, and the bispectrum has dimension $[B_{\Rc}]= [H^4/k^6]$.) }

To end this section, we would like to make several comments.
    \itm{
        \item In this work we investigate multi-field models with Lagrangian of the form (\ref{specific_model}), which are general function
        of two independent contractions $X=X^I_I$ and $Y=X^I_JX^J_I$. One may expect at first that the parameter space for the
        trispectra will become more complicated. However, as we have shown explicitly in this
        work, the final (leading-order) trispectrum for $\mathcal{R}$ is controlled
        by six ``parameters": $\epsilon$, $\ca$, $\ce$, $\lambda$, $\Pi$ and
        $T_{\Rc\Sc}$. Especially, the non-linear structure of $P(X,Y,\phi^I)$
        affect the final trispectrum through $\ca$, $\ce$, $\lambda$ and $\Pi$, which are specific combinations
        of derivative of $P$ with respect to $X$ and $Y$ (see (\ref{def_lambda_Pi}) and (\ref{def_ca_ce})). On the
        other hand, this fact would make it more difficult to determine the structure of a multi-field model from observations, since the
        functional forms of $P$ are highly degenerate with respect to the
        parameter space of trispectra{\footnote{This is very different from single-field case,
        where the large trispectra are expected
        determined by three parameters $c_s$, $\lambda$ and $\Pi$ (or their functions) which are combinations of three derivatives $P_{,XX}$, $P_{,XXX}$ and $P_{,XXXX}$. Thus in principle one may strict the functional form $P(X)$ fairly when $c_s$, $\lambda$ and $\Pi$ have been known.}}.

        \item In the limit of small speed sounds ($\ca \ll 1$, $\ce \ll
        1$), at the leading-order, there are three types of ``contact four-point interaction
        vertices" according to the ``derivatives": vertices with four temporal derivatives, vertices with
        two temporal derivatives and two spatial derivative,
        vertices
        with four spatial derivatives (as illustrated in e.g. Fig.\ref{fig_4adibatic}). This is exactly the same as
        in general single-field models, where three types of contact interaction vertices  correspond to
        three different ``$c_s$-independent" shape functions. However, the subtlety is that, although in
        multi-field case the interaction vertices can still be grouped into these three types as in single-field
        case, since there are more fields and more speeds of sound, parameters arising
from the non-linear structure of the theory enter into the
definition of shape functions, as in (\ref{new_A1}), (\ref{new_A2})
and (\ref{new_A3}). Thus, in general there is no hope to abstract
pure momentum-dependent shape functions without involving
$\epsilon$, $\ca$, $\ce$, $\lambda$, $\Pi$ and $T_{\Rc\Sc}$, and
thus the so-called shape functions in general take the form
    \[
        \mathcal{A}(\vk_1,\vk_2,\vk_3,\vk_4;\epsilon,\ca,\ce,\lambda,\Pi,T_{\Rc\Sc})
        \,.
    \]
    This fact makes the analysis of the shapes of non-Gaussianities in multi-field models
    much more complicated.
}

\section{Characterizing the Trispectrum}

For the power spectrum, it is easy to plot a curve for $P(k)$ which
is a function of a single variable $k$. However, for higher-order
correlation functions, or their fourier transformations bispectrum,
trispectrum, etc, it is not easy to abstract simple quantities to
characterize them. While from the point of view of comparing
theories with observations, this is obviously one of the most
important aspects in higher-order statistics of cosmological
perturbations.

In this section, first we analyze the ``parameter space" for the
trispectrum or general higher-order correlation functions. Base on
the discuss on ``inequivalent momentum configurations", we plot
various ``shape" diagrams which characterize the trispectrum on a
2-dimensional subspace of its whole parametric space. These ``shape"
diagrams are efficient for characterizing the trispectrum (or
higher-order correlation functions) and are also convenient for
visualization. In the end of this section, we map the trispectrum to
 real numbers, which can be seen as the analogue of
non-linear parameters $\tau_{\textrm{NL}}$ and $g_{\textrm{NL}}$ in
the literatures.

\subsection{Parameter Space for the Trispectrum}

In general, the four-point correlation function, or its Fourier
transformation, the trispectrum, is a function of four 3D momenta:
$T\lrp{\vk_1,\vk_2,\vk_3,\vk_4}$. Let us first analyze the
``dimension" of the parameter space of
$T\lrp{\vk_1,\vk_2,\vk_3,\vk_4}$, i.e., the number of independent
degrees of freedom in order to identify a ``inequivalent momentum
configuration" (in other words, the dimension of ``equivalent class"
of momentum configurations). At first, we need 12 real numbers to
specify a set of four 3D momenta $\{ \vk_1, \vk_2, \vk_3, \vk_4 \}$.
However, the 3D momentum conservation ($\vk_1 + \vk_2 + \vk_3 +
\vk_4=0$) eliminate 3 of them, and since the cosmological
perturbation is assumed to be statistically homogeneous and
isotropic{\footnote{See \cite{Gao:2009vi} for a recent investigation
on models which break statistical isotropy, where the quantum
fluctuations are generally statistical anisotropy, and the power
spectrum not only depends on $k=|\vk|$, but also depends on the
orientation of $\vk$.}}, we have 3 dimensional rotation symmetry
SO(3) which eliminates another 3 of them (that is, we call two
momentum configurations are ``equivalent" if they can be related by
SO(3) transformation{\footnote{Of course there have some discrete
symmetry, e.g. permutation symmetry, but this will now affect the
dimension of continuous parameter space.}}). Thus, the dimension of
parameter space for trispectrum is 6. This argument can be
generalized to arbitrary higher-order correlation functions. In
general we need
    \eq{
        3n-6 \,,\qquad (n\geq 3)
    }
real parameters to specify a momentum configuration for a $n$-point
function. While case for $n=2$ is trivial, it is well-known that the
power spectrum is a function of one parameter $k \equiv |\vk|$. For
example, the three-point function or its fourier transformation
bispectrum takes the form $B(k_1,k_2,k_3)$, which depends only on
the modulus of three momenta, which are the nature choices of
parameters for the bispectrum. In the case of trispectrum, things
become much more involved, since we need 6 parameters.
    \itm{
    \item One possible choice is to choose a set of 6 parameters
$\{k_1, k_2, k_3, k_4, k_{12}, k_{23} \}$ with $k_{12} \equiv |\vk_1
+ \vk_2|$ and $k_{23} \equiv |\vk_2 + \vk_3|$. A pictorial
presentation is depicted in fig. \ref{fig_tetra}.
\begin{figure}
    \centering
    \begin{minipage}{0.9\textwidth}
    \centering
        \begin{minipage}{0.45\textwidth}
        \centering
            \begin{minipage}{0.9\textwidth}
            \centering
            \includegraphics[width=6cm]{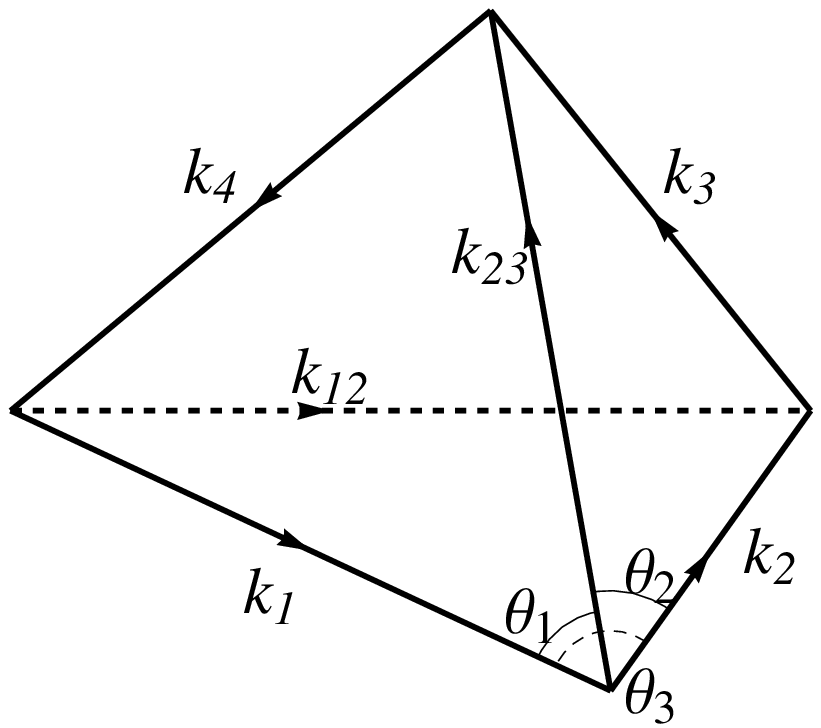}
            \caption{Pictorial representation of sets of six parameters $\{k_1, k_2, k_3, k_4,
k_{12}, k_{23} \}$ or
$\{k_1,k_2,k_{12},\theta_1,\theta_2,\theta_3\}$ to specify an
inequivalent 3D momentum configuration for trispectrum.}
            \label{fig_tetra}
            \end{minipage}
        \end{minipage}
    \begin{minipage}{0.45\textwidth}
    \centering
        \begin{minipage}{0.9\textwidth}
        \centering
        \includegraphics[width=5cm]{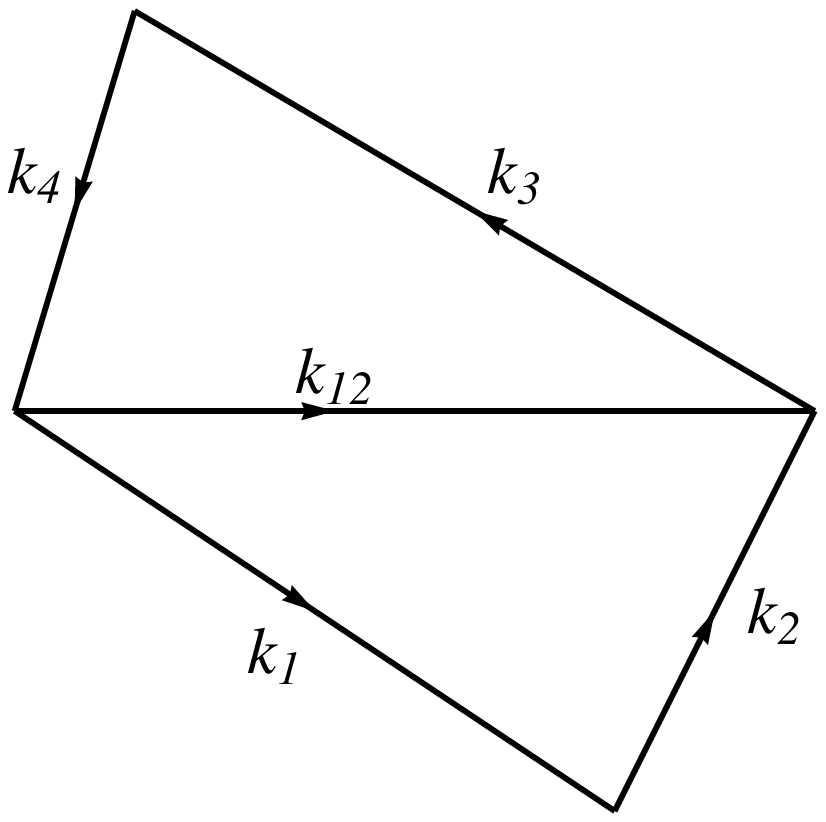}
         \caption{Pictorial representation of a sef of five parameters $\{k_1, k_2, k_3, k_4,
k_{12} \}$ to specify an inequivalent planar momentum configuration
for trispectrum.}
            \label{fig_planar}
            \end{minipage}
        \end{minipage}
         \end{minipage}
        \end{figure}
At this point, it is useful to write down other four
$k_{ij}\equiv|\vk_i + \vk_j|$ in terms of $\{k_1, k_2, k_3, k_4,
k_{12}, k_{23} \}$ for later convenience{\footnote{Remember that
since momentum conservation is: $\vk_1 + \vk_2 + \vk_3 + \vk_4=0$,
it immediately follows $k_{13}=|\vk_1+\vk_3| = |\vk_2 + \vk_4| =
k_{24}$.}}
    \ea{{\label{other_4momenta}}
        k_{13} &\equiv k_{24} = \sqrt{ k_1^2+ k_2^2 + k_3^2 + k_4^2 -k_{12}^2 - k_{23}^2 } \,,\\
        k_{14} &\equiv k_{23}  \,,\qquad\qquad k_{34} \equiv  k_{12} \,.
    }
    Obviously, these six momenta cannot take all real values
    simultaneously. Actually they should satisfy the so-called
    ``triangle inequalities" (remember we have (\ref{other_4momenta})):
        \eq{
            k_i + k_j \geq k_{ij}\,,\qquad k_i + k_{ij} \geq
            k_j\,,\qquad
            (i,j=1,2,3,4\quad \textrm{and} \quad i\neq j)\,.
        }
    \item Another possible parameterization is to choose three momenta and three angles: $\{
k_1,k_2,k_{23}, \theta_1,\theta_2,\theta_3 \}$ (see also
fig.\ref{fig_tetra}). Similarly, the three angles
$\theta_1,\theta_2,\theta_3$ should satisfy:
    \ea{
        \theta_i + \theta_j &\geq \theta_k \,,\qquad
        (i,j,k=1,2,3\,\quad \textrm{and}\quad i\neq j\neq k)\,,\\
        \theta_1+\theta_2+\theta_3 &\leq 2\pi \,.
    }
}

There can be simplifications from the point of view of comparing
theory to observations. Since the Cosmic Microwave Background is
essentially a two-dimensional statistical field, there is actually
one constraint among the above six momentum parameters. Most
significantly, the ``\emph{planar momentum configuration}" is of
special importance, {\footnote{Future LSS (Large-Scale Structure)
experiments will bring us valuable information on 3D-configuration
of the trispectrum.}}, since in practice one average over numbers of
small regions in the CMB, which are approximately planar. Thus, in
the two-dimensional case, we need (2 dof from momentum conservation
and 1 from SO(2) rotation symmetry)
    \eq{
        2n-3
    }
parameters to characterize a planar momentum
configuration for $n$-point function. For example, for three-point
function (or bispectrum), the momentum configuration is always a
 triangle, thus there are $3$ independent parameters to
determine a configuration of bispectrum. While we need $5$
independent parameters to specify a planar momentum configuration
for trispectrum. How to choose these five parameters properly is
somewhat tricky. In this work, we choose the five parameters as
$\{k_1, k_2, k_3, k_4, k_{12} \}$, which is depicted in
fig.\ref{fig_planar}. Moreover, these four momenta must satisfy a
``planar condition", this condition can be expressed as writing
$k_{23}$ in terms of other 5 momenta:
    \ea{
        k_{23} &\equiv \sqrt{ k_1^2 + k_4^2 - 2k_1 k_4 \cos(\alpha + \beta)
        } \,,\\
        \textrm{with} \qquad \cos \alpha &\equiv  \frac{k_4^2 + k_{12}^2 - k_3^2}{2 k_4
        k_{12}} \,,\qquad \cos\beta \equiv \frac{k_1^2 + k_{12}^2 - k_2^2}{2 k_1
        k_{12}} \,.
    }

\subsection{Shape of the Trispectrum}

In this section we plot the shape diagrams of the trispectrum, i.e.
two dimensional surfaces which capture parts of the properties of
the trispectrum{\footnote{Since the momentum configuration space for
the trispectrum is high-dimensional (6 for 3D configuration, 5 for
2D configuration), the proper visualization of the trispectrum is
still a challenge. The traditional ``shape diagrams" are
two-dimensional surfaces which captures parts of the properties of
the trispectrum, i.e. the projection onto a two-dimensional
subspace.}}. In this work we plot the shape diagrams with  one 3D
momentum configuration and one 2D momentum configuration.

\begin{figure}[h]
    \centering
    \begin{minipage}{1\textwidth}
    \centering
        \begin{minipage}{0.45\textwidth}
            \centering
            \includegraphics[width=7cm]{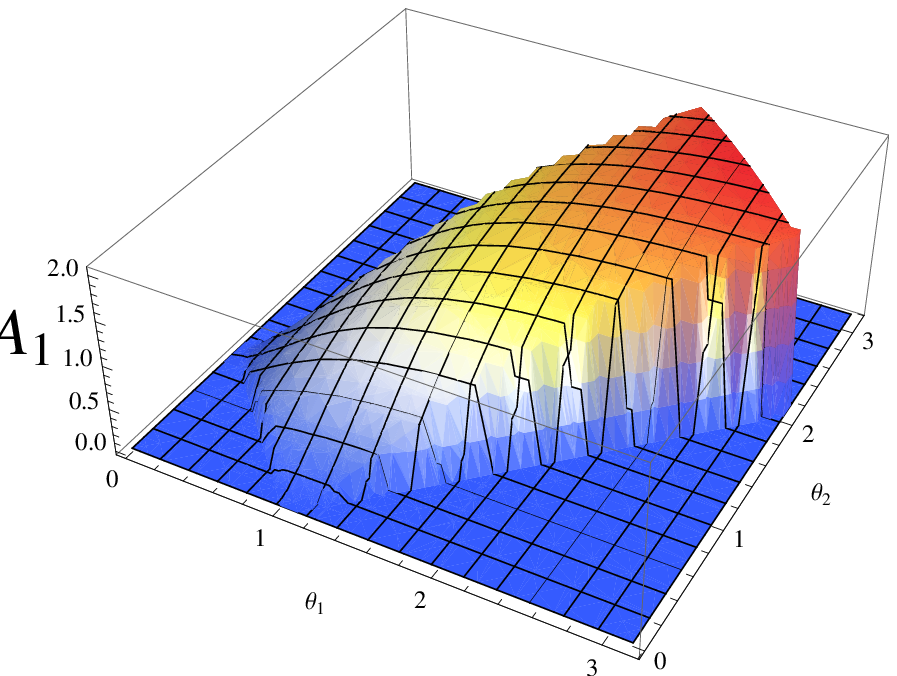}
        \end{minipage}
        \begin{minipage}{0.45\textwidth}
            \centering
            \includegraphics[width=7cm]{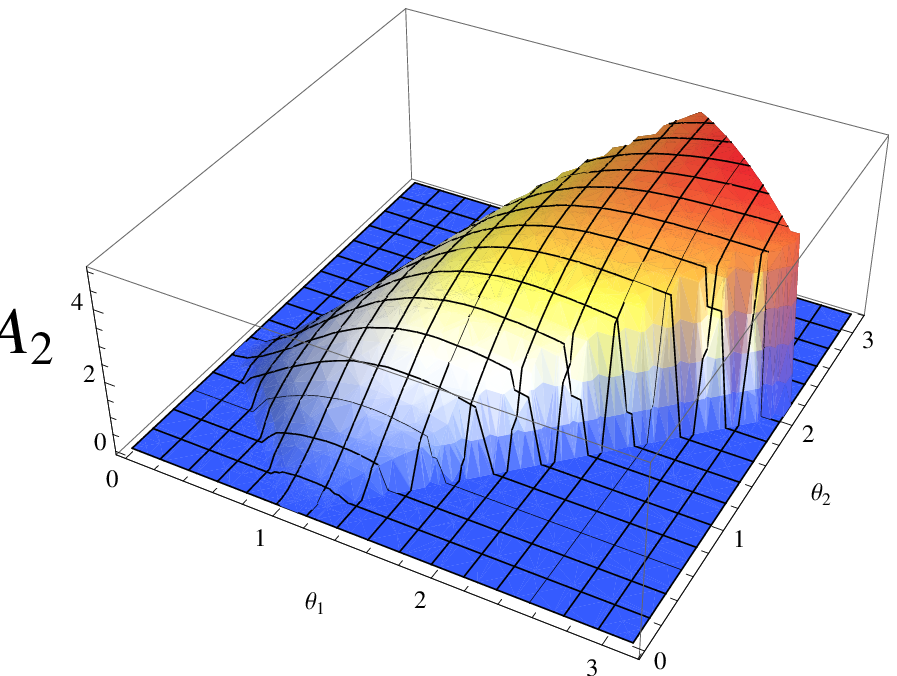}
        \end{minipage}
        \begin{minipage}{0.45\textwidth}
            \centering
            \includegraphics[width=7cm]{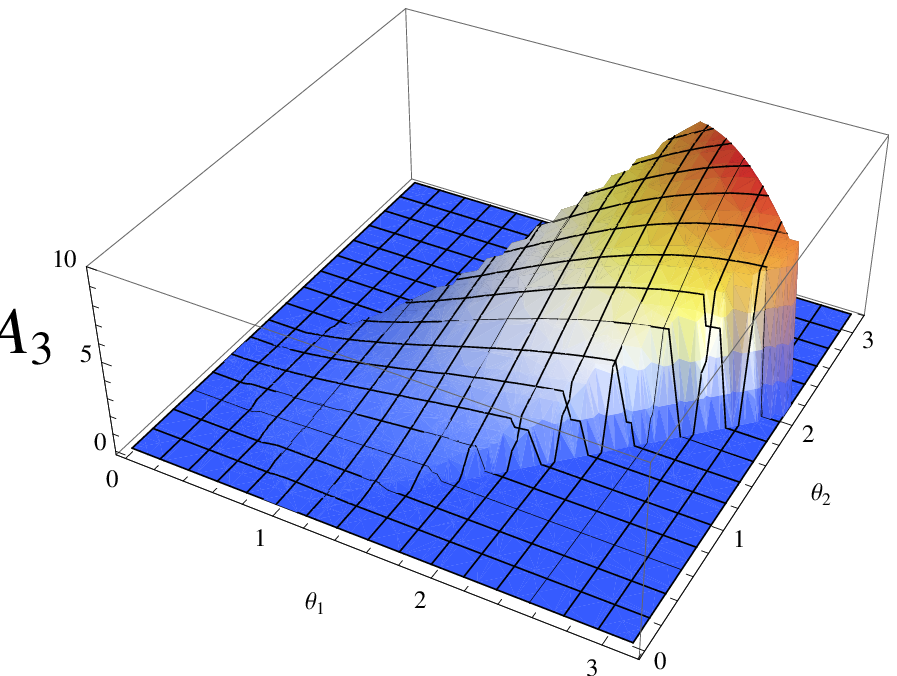}
        \end{minipage}
    \end{minipage}
    \caption{Shapes of $\mathcal{A}_1$, $\mathcal{A}_2$ and $\mathcal{A}_3$ with momentum configuration $k_1=k_2=k_{12}=k_{23}$ as function of $\theta_1$ and $\theta_2$. In this group of diagrams, parameters are chosen as $\ca=\ce=0.1$, $\epsilon=0.01$, $\lambda=\Pi=100$ and $T_{\Rc\Sc}=1$.}
    \label{fig_3D_shape}
    \end{figure}

\begin{figure}[h]
    \centering
    \begin{minipage}{1\textwidth}
    \centering
        \begin{minipage}{0.45\textwidth}
            \centering
            \includegraphics[width=7cm]{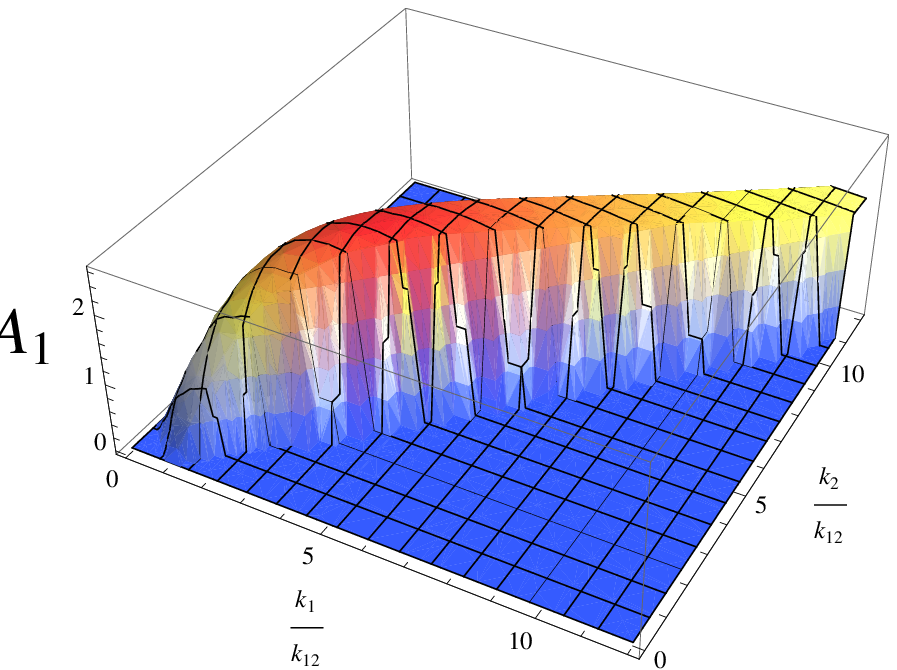}
        \end{minipage}
        \begin{minipage}{0.45\textwidth}
            \centering
            \includegraphics[width=7cm]{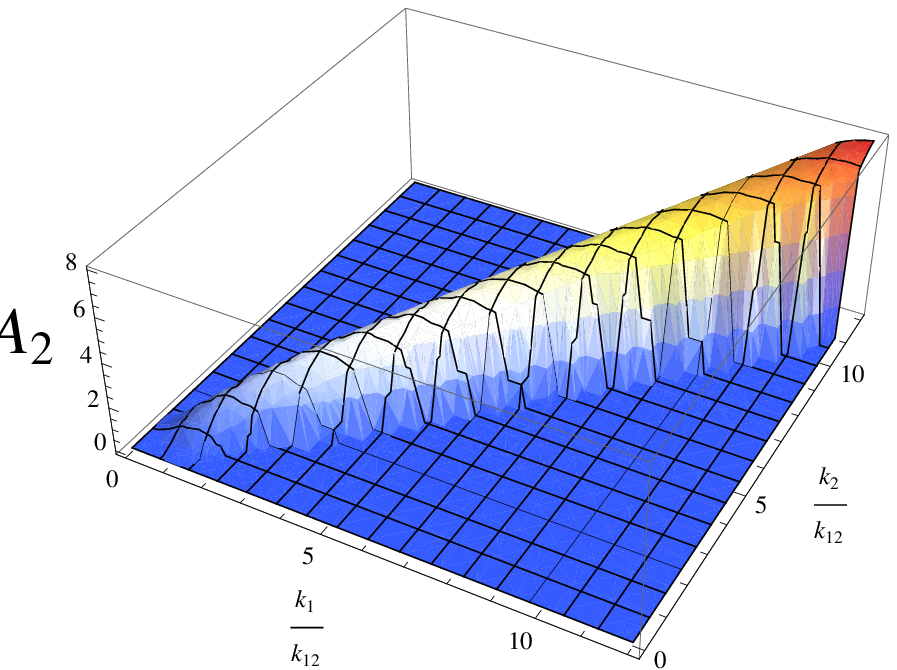}
        \end{minipage}
        \begin{minipage}{0.45\textwidth}
            \centering
            \includegraphics[width=7cm]{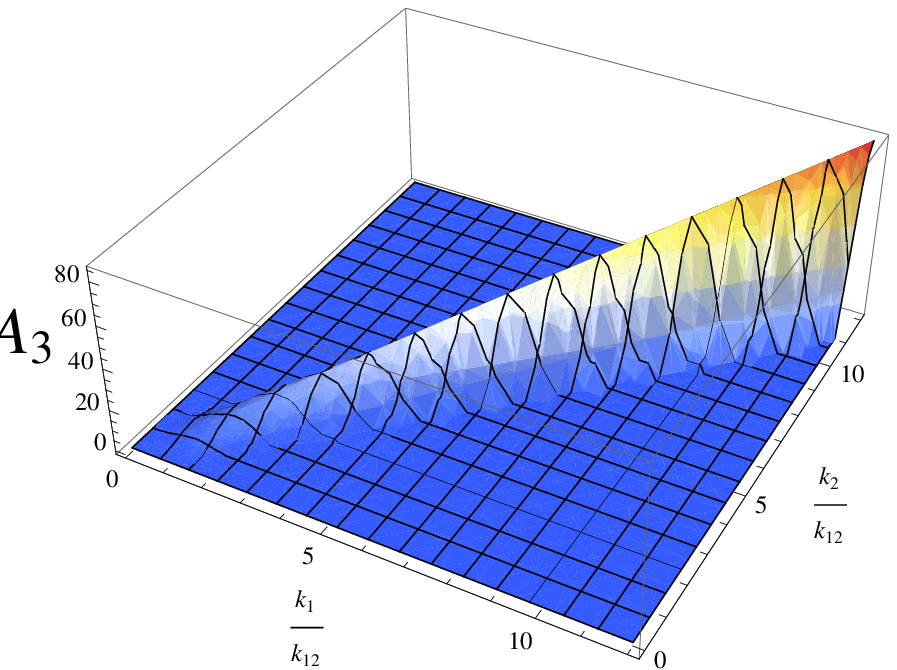}
        \end{minipage}
    \end{minipage}
    \caption{Shapes of $\mathcal{A}_1$, $\mathcal{A}_2$ and $\mathcal{A}_3$ with 2D momentum configuration $k_1=k_2=k_{12}$ as function of $k_1$ and $k_2$. In this group of diagrams, parameters are chosen as $\ca=\ce=0.1$, $\epsilon=0.01$, $\lambda=\Pi=100$ and $T_{\Rc\Sc}=1$.}
    \label{fig_2D_shape}
    \end{figure}

\itm{
    \item For 3D momentum configurations, we would like to choose $k_1 =
k_2 = k_{12} = k_{23}$. Thus we are left with two free parameters
which one may choose as $k_3$ and $k_4$. Actually in this case, it
is more convenient to choose two angles: $\theta_1$ and $\theta_2$,
which satisfy the inequalities $\frac{\pi}{3} \leq \theta_1+\theta_2
\leq \frac{5\pi}{3}$ and $|\theta_1 - \theta_2| \leq \frac{\pi}{3}$.
Note that we have
    \[
        k_3 = \sqrt{ k_2^2 + k_{23}^2 - 2 k_2 k_{23} \cos\theta_2
        }\,,\qquad\qquad k_4 = \sqrt{ k_1^2 + k_{23}^2 - 2k_1 k_{23} \cos\theta_1
        }\,.
    \]
The shape diagrams of $\mathcal{A}_1$, $\mathcal{A}_2$ and
$\mathcal{A}_3$ are depicted in fig.\ref{fig_3D_shape}.

    \item For planar momentum configurations (see Fig.\ref{fig_planar}), we
consider $k_3 = k_4 =
        k_{12}$, thus the parameter space becomes $\{k_1,k_2\}$. The
        shapes are depicted in
        fig.\ref{fig_2D_shape}.
        }

\subsection{Non-linear Parameters}

In practice it is also convenient to define some so-called
non-linear parameters, which characterize the typical or overall
amplitudes of non-Gaussianities{\footnote{Mathematically, this is
essentially to map $T(\vk_1,\vk_2,\vk_3,\vk_4)$ to a real number.
The ``inequivalent momenta configuration" space $\mathcal{M}$ for
the trispectrum $T(\mathcal{M})$ is 6-dimensional. Traditional
definitions of non-linear parameters correspond to choosing one
specific point $p \in \mathcal{M}$ (i.e. one specific momenta
configuration), and defining the non-linear paramter(s) proportional
to $T|_{p\in\mathcal{M}} \in \mathbb{R}$.}}. For bispectrum which is
defined as
    \eq{
        \lrab{ \mathcal{R}(\vk_1)\mathcal{R}(\vk_2)\mathcal{R}(\vk_3)}
        = (2\pi)^3 \delta^3(\vk_{123}) B(k_1,k_2,k_3)
        \,,
    }
the non-linear parameter $f_{\textrm{NL}}$ is usually defined as
    \eq{{\label{f_NL}}
         B(k_1,k_2,k_3) \equiv \frac{6}{5} f_{\textrm{NL}}(k_1,k_2,k_3) \lrsb{ P(k_1)P(k_2) + P(k_2)P(k_3) + P(k_3)P(k_1)
         } \,.
    }
Note that in general $f_{\textrm{NL}}(k_1,k_2,k_3)$ is a function of
three momenta. (\ref{f_NL}) is motivated from the ``local"-type
bispectrum, which arises from interaction of the form as local
product in real space:
    \eq{{\label{local_bi}}
        \Rc = \Rc_\textrm{g} + \frac{3}{5}
        f_{\textrm{NL}}^{\textrm{local}} \lrp{ \Rc_{\textrm{g}}^2 - \lrab{\Rc_{\textrm{g}}}^2
        } \,,
    }
where $\Rc_{\textrm{g}}$ is Gaussian. In this case,
$f_{\textrm{NL}}$ is a pure real number, and the bispectrum is
maximized in the limit of one of the three momenta going to zero.
Although interaction of the form (\ref{local_bi}) is simple in real
space, it is difficult to generate large local bispectrum in
realistic models. Actually this form of non-Gaussianities are mostly
produced by nonlinear gravitational evolution subsequent to the
horizon-crossing \cite{Babich:2004gb,Lyth:2005fi}.

For trispectrum as defined in (\ref{trispectrum_def}), the case is
much more complicated. One possible parameterization is
    \eq{{\label{tau_NL_g_NL}}
        T(\vk_1,\vk_2,\vk_3,\vk_4) \equiv \tau_{\textrm{NL}} \lrsb{ P(k_{13})P(k_3)P(k_4) + \textrm{11 perms}
        } + \frac{54}{25} g_{\textrm{NL}} \lrsb{ P(k_1)P(k_2)P(k_3) + \textrm{3 perms}
        } \,,
    }
where $k_{ij} \equiv |\vk_i + \vk_j|$. This parameterization of
trispectrum is also motivated by the ``local-product-type"
interaction in real space:
    \eq{{\label{local_tri}}
        \Rc \equiv \Rc_{\textrm{g}} + \frac{1}{2} \lrp{ \tau_{\textrm{NL}}^{\textrm{local}}
        }^{\frac{1}{2}} \lrp{ \Rc^2_{\textrm{g}} - \lrab{ \Rc_{\textrm{g}} }^2
        } + \frac{9}{25} g_{\textrm{NL}}^{\textrm{local}} \lrp{ \Rc^3_{\textrm{g}} -3 \Rc_{\textrm{g}} \lrab{ \Rc^2_{\textrm{g}} }
        } \,.
    }
Actually, (\ref{tau_NL_g_NL}) itself has no prior convenience for
the parameterization of trispectrum in general context. This is
justified by the fact that, in models with non-canonical kinetic
terms, especially with small speed(s) of sound, the leading-order
non-Gaussianities are originated from the ``derivative-coupled"-type
interactions, which is obviously not real space
``local-product"-type interactions described as (\ref{local_bi}) and
(\ref{local_tri}).

Actually, there is a general definition of non-linear parameters for
the trispectrum. From (\ref{trispectrum_R}) and the discussion in
sec.\ref{sec_tri}, the trispectrum has dimension $\frac{H^6}{k^9}$,
thus one may conveniently define a dimensionless non-linear
parameter $\tau_{\textrm{NL}}$ by
    \eq{{\label{nl_def_general}}
        T_{\Rc}(\vk_1,\vk_2,\vk_3,\vk_4) \equiv (2\pi)^6
        \mathcal{P}_{\Rc}^3 \frac{\tau_{\textrm{NL}}(\vk_1,\vk_2,\vk_3,\vk_4)}{
        F(\vk_1,\vk_2,\vk_3,\vk_4)} \,,
    }
where $\mathcal{P}_{\Rc}$ is the power spectrum of curvature
perturbation on large scales which is given in (\ref{spectra_ls})
and $F(\vk_1,\vk_2,\vk_3,\vk_4)$ is some function of momenta which
has momentum dimension $k^9$. Remember that in general the so-called
non-linear parameters $\tau_{\textrm{NL}}$ are always functions of
momenta.

As addressed before, the non-linear parameter is essentially a
characterization of the typical or overall amplitude of
non-Gaussianities. Thus one may freely choose convenient
$F(\vk_1,\vk_2,\vk_3,\vk_4)$ and momenta configurations to abstract
real numbers for various purpose. In this work, in order to get a
glance of the amplitude of non-Gaussianities from trispectrum, we
abstract one real number $\tau^{\textrm{rth}}_{\textrm{NL}}$ from
the trispectrum, which are defined
respectively{\footnote{$\tau^{\textrm{rth}}_{\textrm{NL}}$ has also
been used in a recent investigation of the trispectrum in general
single field models \cite{Chen:2009bc}.}}:
    \ea{{\label{nl_para}}
        \left. T_{\Rc}(\vk_1,\vk_2,\vk_3,\vk_4)\right|_{\textrm{regular tetrahedron}}
        &\equiv   (2\pi)^6
        \mathcal{P}_{\Rc}^3 \frac{1}{
        k^9} \, \tau_{\textrm{NL}}^{\textrm{rth}} \,,
    }
in which ``regular tetrahedron" denotes the ``regular tetrahedron"
momenta configuration for trispectrum:
$k_1=k_2=k_3=k_4=k_{12}=k_{23}\equiv k$ (see fig.\ref{fig_tetra}).
Here we conveniently choose $F = k^9$. One can of course define
other convenient non-linear parameters by choosing specific momentum
configurations and $F$ in (\ref{nl_def_general}) for particular
purpose.

As has been discussed in the end of sect.\ref{sec_tri}, in multi-field
case, the shape functions intrinsically involve parameters
like $\ca$ etc. Thus, even the momentum configuration is fixed, the
trispectrum or the corresponding non-linear parameters defined as in
(\ref{nl_para}) is complicated. In this work we give the expression
in the limit $\ca \ll 1$ and $\ce \ll 1$. We have
    \ea{
        \tau_{\textrm{NL}}^{\textrm{rth}} &= \lrp{1 + T^2_{\Rc\Sc} \frac{\ca}{\ce}}^{-3} \lrp{\tau_1 + \tau_2 + \tau_3 }\,,
    }
where
    \ea{
        \tau_1 &= - \frac{3 \ca^2 \left(54 \ca^2 \lambda ^2-H^2 \epsilon  (3 \lambda +10 \Pi )\right)}{512 H^4 \epsilon ^2} - T^2_{\Rc\Sc} \frac{9 \ca  \left(H^2 \epsilon +3 \ca^4 \left(-7 \ca^2+2 \ce^2\right) \lambda \right)}{8 (\ca+\ce)^5 H^2 \epsilon } - T^4_{\Rc\Sc} \frac{9 \ca \left(-13 \ca^2+4 \ce^2\right) }{1024 \ce^5}\,,\\
        \tau_2 &= \frac{13 \left(-H^2 \epsilon + 3 \ca^4 \lambda \right)}{256 \ca^2 H^2 \epsilon} + T^2_{\Rc\Sc} \frac{\ca^3 \left(\ca^2+5 \ca \ce+7 \ce^2\right)  \left(-H^2 \epsilon +3 \ca^4 \lambda \right)}{4 \ce^4 (\ca+\ce)^5 H^2 \epsilon } + T^4_{\Rc\Sc}\frac{13 \ca (\ca^2 - \ce^4) }{256 \ce^7} \,,\\
        \tau_3 &=  \frac{515}{8192 \ca^2}+ T^2_{\Rc\Sc}\frac{\ca \left(-5 \ca^2+ 6 \ce^2\right) \left(5 \ca^4+25 \ca^3 \ce +43 \ca^2 \ce^2+25 \ca \ce^3+5 \ce^4\right) }{64 \ce^4 (\ca+\ce)^5}  - T^4_{\Rc\Sc}\frac{515 \ca^3 \left(\ca^2-\ce^2\right) }{2048 \ce^9}
        \,.
    }

\section{Conclusion and Discussion}

In this work, we studied the inflationary trispectrum in general
multi-field models with scalar field Lagrangian of the form
$P(X^{IJ},\phi^I)$, which is the generalization of multi-field
$k$-inflation and multi-field DBI models. In our general framework,
we expanded the perturbation action up to the fourth-order, and
calculated the four-point correlation functions for adiabatic and
entropy modes, and also the trispectrum of the curvature
perturbation on superhorizon scales.

We have shown that the perturbations for adiabatic and entropy modes
are enhanced by $1/\ca$ and $1/\ce$ respectively, where $\ca$ and
$\ce$ are propagation speeds of adiabatic and entropy modes
respectively and in general $\ca \neq \ce$. In this work we focus on
the trispectrum from four-point interaction vertices. In the
two-field case, we have shown that there are three non-vanishing
four-point correlation functions: $\lrab{\qsg\qsg\qsg\qsg}$,
$\lrab{\qsg\qsg\qs\qs}$ and $\lrab{\qs\qs\qs\qs}$. In the
leading-order, all four-point correlation functions and also the
final trispectrum of the curvature perturbation can be grouped into
three types, which are deformations and permutations of three
fundamental shape functions (\ref{basic_shapes}). These three types
of contributions arise from three types of four-point interaction
vertices. Thus, our result can be seen as the generalization of the
previous results in general single-field models
\cite{Huang:2006eh,Arroja:2008ga,Chen:2009bc,Arroja:2009pd} and also
the very recent investigations in multi-DBI models
\cite{Mizuno:2009cv}.

In single-field models, one can always abstract momentum-dependent
shape functions and put all other parameters as overall pre-factors.
However, as we have seen in three ``generalized shape functions"
(\ref{new_A1})-(\ref{new_A3}), parameters such as $\ca$, $\ce$,
$\lambda$, $\Pi$ and $T_{\Rc\Sc}$ enter the definitions of these
shape functions, and make the momentum dependence of these shape
functions complicated. In this work, after a general discussion on
the parameter-dependence of the trispectrum, we plot two set of
shape diagrams of (\ref{new_A1})-(\ref{new_A3}), including 3D and 2D
momentum configuration with fixed parameters as $\ce$ etc. However,
one expects that there should be better characterization of the
trispectrum and the shape functions. We hope to get back to these
issues in the future.

In this work we focus on the non-Gaussianities arising from the
interactions among quantum fluctuations. While interactions not only
cause non-Gaussianities, they also cause quantum loop-corrections.
By collecting signatures of both quantum loop corrections and
non-Gaussianities, we will obtain a more sensitive test of the
physics of inflation \cite{Gao:2009fx}.

In this work we focus on the primordial non-Gaussianities which are
evaluated around the sound-horizon crossing. While detectable
non-Gaussianities can also be produced when the curvature
perturbation is generated from the entropy perturbation(s) on
superhorizon scales, or at the end of inflation, or during the
complicated reheating process. Since non-canonical kinetic terms can
arise naturally in string theory inspired models, cosmic string
effects should also be considered \cite{Sarangi:2002yt,Feng:2007qk}.

Another issue should be address is that, in this work (and also
previous works on non-Gaussianities in general multi-field models),
the cross-correlation between adiabatic mode and entropy mode around
horizon-crossing is assumed vanishing: $\lrab{\qsg\qs}_{\ast}
\approx 0$. This is good approximation in slow-roll inflation with
canonical kinetic term, which corresponds to $\tilde{\xi}\approx 0$
in (\ref{tilde_xi}) and the background strategy is straight. In this
work we also assume $\tilde{\xi} \approx 0$, while it is interesting
to investigate the case where $\tilde{\xi}$ cannot be neglected and
$\lrab{\qsg\qs}_{\ast}\neq 0$
\cite{Langlois:2005ii,Byrnes:2006fr,Lalak:2007vi}.

\bigskip
{\bf Acknowledgements}

    We would like to thank Qing-Guo Huang, Yi Wang for
    helpful discussions. XG thank Frederico Arroja and Kazuya Koyama for useful correspondence. This work was supported by the NSFC grant
No.10535060/A050207, a NSFC group grant No.10821504 and
Ministry of Science and Technology 973 program under grant No.2007CB815401.

\bigskip
\bigskip
\appendix

\section{A Brief Review of ``In-in" Formalism}{\label{appsec_inin}}

\subsection{Preliminaries}

The ``in-in formalism" (also dubbed as ``Schwinger-Keldysh
formalism", or ``Closed-time path formalism")
\cite{Schwinger:1960qe,Calzetta:1986ey,Jordan:1986ug} is a
perturbative approach for solving the evolution of expectation
values over a finite time interval. It is therefore ideally suited
not only to backgrounds which do not admit an S-matrix description,
such as inflationary backgrounds.

In  the calculation of S-matrix in particle physics,
the goal is to determine the amplitude for a state in the far past
$|\psi\ra$ to become some state $|\psi'\ra$ in the far future,
    \[
        \la \psi' | S |\psi \ra = \la \psi'(+\infty)
        |\psi(-\infty)\ra \,.
    \]
Here, conditions are imposed on the fields at both very early and
very late times. This can be done because that in Minkowski
spacetime, states are assumed to be non-interacting at far past
and at far future, and thus are usually taken to be the free vacuum,
i.e., the vacuum of the free Hamiltonian $H_0$. The free vacuum are
assumed to be in ``one-to-one" correspondence with the true vacuum
of the whole interacting theory, as we adiabatically turn on and
turn off the interactions between $t=-\infty$ and $t=+\infty$.

While the physical situation we are considering here is quite
different. Instead of specifying the asymptotic conditions both in
the far past and far future, we develop a given state \emph{forward}
in time from a specified initial time, which can be chosen as the
beginning of inflation. In the cosmological context, the initial state
is usually chosen as free vacuum, such as Bunch-Davis vacuum, since
at very early times when perturbation modes are deep inside the
Hubble horizon, according to the equivalence principle, the
interaction-picture fields should have the same form as in Minkowski
spacetime.

\subsection{``In" vacuum}

The Hamiltonian can be split into a free part and an interacting
part: $H=H_0+H_{\textrm{i}}$. The time-evolution operator in the
interacting picture is well-known
    \eq{{\label{U_op}}
        U(\eta_2,\eta_1) = \mathrm{T}\exp\left( -i\int_{\eta_1}^{\eta_2} dt' H_{\textrm{i}\I}(\eta')
        \right)\,,
    }
where subscript ``$\I$" denotes interaction-picture quantities,
$\textrm{T}$ is the time-ordering operator. Our present goal is to
relate the interacting vacuum at arbitrary time $|\Omega_\I(t)\ra$
to the free vacuum $|0_\I\ra$ (e.g., Bunch-Davis vacuum). The trick is
standard. First we may expand $|\Omega_\I(\eta)\ra$ in terms of
eigenstates of free Hamiltonian $H_0$, $|\Omega_\I(\eta)\ra = \sum_n
|n_\I\ra\, \la n_\I| \Omega_\I(\eta)\ra$, then we evolve
$|\Omega_\I(\eta)\ra$ by using (\ref{U_op})
    \eq{{\label{vacuum_evo}}
        |\Omega_\I(\eta_2)\ra = U(\eta_2,\eta_1)
        |\Omega_\I(\eta_1)\ra = |0_\I\ra\,
        \la 0_\I| \Omega_\I\ra + \sum_{n\geq1} e^{+iE_n(\eta_2 - \eta_1)}\, |n_\I \ra\,
        \la n_\I| \Omega_\I(\eta_1)\ra \,.
    }
From (\ref{vacuum_evo}), we immediately see that, if we choose
$\eta_2 = -\infty(1-i\epsilon)$, all excited states in
(\ref{vacuum_evo}) are suppressed. Thus we relate interacting vacuum
at $\eta= -\infty(1-i\epsilon)$ to the free vacuum $|0\ra$ as
    \eq{
        |\Omega_\I(-\infty (1 - i\epsilon) )\ra = |0_\I \ra\,
        \la 0_\I| \Omega_\I \ra
    }
Thus, the interacting vacuum at arbitrary time $\eta$ is given by
    \ea{
        |\textrm{VAC},\textrm{in} \ra &\equiv |\Omega_\I(\eta)\ra =  U(\eta,-\infty(1-i\epsilon))  |\Omega_\I(-\infty(1-i\epsilon))\ra \\
        &= \mathrm{T}\exp\left( -i\int^{\eta}_{-\infty(1-i\epsilon)} d\eta'\, H_{\textrm{i}\I}(\eta')
        \right) |0_\I \ra\,
        \la 0_\I| \Omega_\I \ra \,.
    }

\subsection{Expectation values in ``in-in" formalism }

The expectation value of operator $\hat{\mathcal{O}}(\eta)$ at
arbitrary time $\eta$ is evaluated as
    \ea{{\label{in-in_ev}}
        \la \hat{\mathcal{O}}(\eta) \ra &\equiv \frac{ \la \textrm{VAC},\textrm{in} | \hat{\mathcal{O}}(\eta) |\textrm{VAC},\textrm{in}
        \ra }{ \la \textrm{VAC},\textrm{in} |\textrm{VAC},\textrm{in}
        \ra } \\
        &= \soev{ 0_\I }{
        \bar{ \mathrm{T}} \exp\left( i\int^{\eta}_{-\infty{ (1 + i\epsilon)} } d\eta' H_{1\I}(\eta')
        \right) \, \hat{\mathcal{O}}_\I(\eta) \, \mathrm{T}\exp\left( -i\int^{\eta}_{-\infty {( 1-i\epsilon)} } d\eta' H_{\textrm{i}\I}(\eta')
        \right) }{0_\I } \,,
    }
where $\bar{\textrm{T}}$ is the anti-time-ordering operator.

\begin{figure}[h]
    \centering
    \begin{minipage}{0.7\textwidth}
    \centering
        \includegraphics[width=8cm]{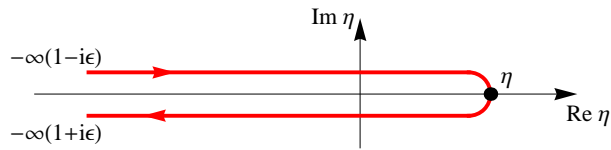}
    \end{minipage}
    \caption{Closed-time path in ``in-in" formalism.}
    \label{fig2_ctp}
    \end{figure}

For simplicity, we denote \eq{
    -\infty(1-i\epsilon) \equiv -\infty^+\,,\qquad\qquad -\infty(1+i\epsilon) \equiv
    -\infty^- \,,
    }
 since, e.g., $-\infty^+$
has a positive imaginary part. Now let us focus on the time-order in
(\ref{in-in_ev}). In standard S-matrix calculations, operators
between $\la 0|$ and $|0\ra$ are automatically time-ordered. While
in (\ref{in-in_ev}), from right to left, time starts from
infinite past, or $-\infty^+$ precisely, to some arbitrary time
$\eta$ when the expectation value is evaluated, then back to
$-\infty^-$ again. This time-contour, which is shown in
Fig.\ref{fig2_ctp}, forms a closed-time path, so ``in-in" formalism
is sometimes called ``closed-time path" (CTP) formalism.

\subsection{Perturbation theory}

The starting point of perturbation theory is the free theory
two-point correlation functions. In canonical quantization
procedure, we write a scalar field as
    \eq{
        \phi_{\vk}(\eta) = u(k,\eta) a_{\vk} + u^{\ast}(k,\eta)
        a^{\dag}_{-\vk} \,,
    }
where $u(k,\eta)$ is the mode function for $\phi_{\vk}(\eta)$ (in
practice, $u_k(\eta)$ and $u^{\ast}_k(\eta)$ are two
linear-independent solutions of equation of motion for
$\phi_{\vk}(\eta)$, which are Wroskian normalized and satisfy some
initial or asymptotic conditions ).

The free two-point function takes the form
    \eq{{\label{2pf}}
        \soev{0}{ \phi_{\vk_1}(\eta_1) \phi_{\vk_2}(\eta_2) }{0}
        \equiv (2\pi)^3 \delta^3(\vk_1+\vk_2)  G_{k_1}(\eta_1,\eta_2)
        \,,
    }
with
    \eq{{\label{free_gf}}
        G_{k_1}(\eta_1,\eta_2) \equiv u_{k_1}(\eta_1)
        u^{\ast}_{k_1}(\eta_2) \,.
    }
In this work, we take (\ref{2pf}) and (\ref{free_gf}) as the
starting point.

Now Taylor expansion of (\ref{in-in_ev}) gives
    \itm{
        \item 0th-order
            \eq{{\label{ev_0th}}
                \lrab{ \hat{\mathcal{O}}(\eta) }^{(0)} = \la 0_\I | \hat{\mathcal{O}}_\I(\eta) |0_\I\ra \,.
                }
        \item 1st-order (one interaction vertex)
            \eq{{\label{app_ev_1st}}
                \lrab{ \hat{\mathcal{O}}(\eta) }^{(1)} = 2\, \textrm{Re} \lrsb{ -i \, \int^{\eta}_{{ -\infty^+} } d\eta'\,
         \soev{0_\I}{ \hat{\mathcal{O}}_\I(\eta)\, H_{\textrm{i}\I}(\eta')  }{0_\I}
            }\,.
            }
         \item 2nd-order (two interaction vertices)
            \eq{{\label{ev_2nd}}
            \begin{aligned}
                \lrab{ \hat{\mathcal{O}}(\eta) }^{(2)} &= - 2\, \textrm{Re} \lrsb{ \int^{\eta}_{{ -\infty^+} } d\eta'\, \int^{\eta'}_{{ -\infty^+}
        } d\eta''\, \soev{0_\I}{ \hat{\mathcal{O}}_\I(\eta)\,  H_{\textrm{i}\I}(\eta')\, H_{\textrm{i}\I}(\eta'') }{0_\I} } \\
        &\qquad\qquad + \int^{\eta}_{{ -\infty^-} } d\eta' \int^{\eta}_{{ -\infty^+} }
        d\eta''\, \soev{0_\I}{ H_{\textrm{i}\I}(\eta') \, \hat{\mathcal{O}}_\I(\eta) \,
         H_{\textrm{i}\I}(\eta'') }{0_\I} \,.
         \end{aligned}
            }
    }
Here in this work, since we are considering four-point correlation
functions from four-point vertices, (\ref{app_ev_1st}) is needed.

\section{Various Expansion Quantities and Useful Relations}\label{app_stacks}

The exact form
    \ea{
        P_4 &= P_{,\ab{IJ}} X^{IJ}_4 + \frac{1}{2 } \lrsb{ P_{,\ab{IJ}\ab{KL}} \lrp{ X^{IJ}_2 X^{KL}_2 + 2\, X^{(IJ}_1
        X^{KL)}_3} + 2\, P_{,\ab{IJ}K} Q^K X^{IJ}_3 } \\
        &\qquad + \left[ \frac{1}{6} P_{,\ab{IJ} \ab{KL}\ab{MN}} \lrp{ X^{IJ}_2 X^{KL}_1 X^{MN}_1 + X^{KL}_2 X^{MN}_1 X^{IJ}_1  +
        X^{MN}_2 X^{IJ}_1 X^{KL}_1 } \right.\\
        &\qquad\qquad \qquad \left. +  P_{,\ab{IJ}\ab{KL}M} Q^M\, X^{(IJ}_1 X^{KL)}_2 + \frac{1}{2} P_{,\ab{IJ}KL}Q^KQ^L\,
        X^{IJ}_2 \right] \\
        &\qquad + \frac{1}{24} \left\{  P_{,\ab{IJ} \ab{KL} \ab{MN} \ab{PQ} }\, ( X^{IJ}_1 X^{KL}_1 X^{MN}_1 X^{PQ}_1 ) + 4\, P_{,\ab{IJ} \ab{KL} \ab{MN} P}  Q^P
    (X^{IJ}_1 X^{KL}_1 X^{MN}_1)
    \right. \\
    &\qquad\qquad \left. + 6\, P_{,\ab{IJ} \ab{KL} MN} Q^MQ^N (X^{IJ}_1 X^{KL}_1) + 4\, P_{,\ab{IJ} KLM} Q^KQ^LQ^M \, X^{IJ}_1 +
    P_{,IJKL}Q^IQ^JQ^KQ^L \right\} \,,
    }
    \ea{
        P_3 &= P_{,\ab{IJ}} X^{IJ}_3 + \lrsb{ P_{,\ab{IJ}\ab{KL}} \,  X^{(IJ}_1 X^{KL)}_2 + P_{,\ab{IJ}K} Q^K
    X^{IJ}_2} \\
    &\qquad + \frac{1}{6} P_{,\ab{IJ} \ab{KL}\ab{MN}} X^{IJ}_1 X^{KL}_1 X^{MN}_1 \\
    &\qquad + \frac{1}{2} P_{,\ab{IJ}\ab{KL}M} Q^M X^{IJ}_1 X^{KL}_1 + \frac{1}{2} P_{,\ab{IJ}KL}Q^KQ^L X^{IJ}_1 + \frac{1}{6}
    P_{,IJK}Q^IQ^JQ^K \,,
    }
and
    \ea{
        P_2 &= P_{,\ab{IJ}} \lrcb{ -\frac{1}{2a^2} \rd_i Q^I \rd_i Q^J  + \frac{1}{2} \lrsb{ \lrp{\doq^I \doq^J - 2\, \dop_0^{(I} N^i_{1} \rd_i Q^{J)}} - 4\, \alpha_1 \, \dop_0^{(I} \doq^{J)} +
        \lrp{ 3 \alpha _1^2 - 2 \alpha_2}\, \dop_0^I \dop_0^J } } \\
        &\qquad\qquad +  \frac{1}{2}\left[ P_{,\ab{IJ}\ab{KL}} \lrp{ \dop_0^{(I} \doq^{J)}  - \alpha_1\, \dop_0^I \dop_0^J } \lrp{ \dop_0^{(K} \doq^{L)}  - \alpha_1\, \dop_0^K \dop_0^L } \right. \\
        &\qquad\qquad \left. + 2\, P_{,\ab{IJ}K} Q^K \lrp{ \dop_0^{(I} \doq^{J)}  - \alpha_1\, \dop_0^I \dop_0^J } + P_{,IJ}
        Q^IQ^J \right] \,.
        }

    \ea{
        X^{IJ}_4 &=  \frac{1}{2}  \lrp{ \rd^i\beta_1 \rd_i Q^I\, \rd^i\beta_1 \rd_i Q^J - 2\, \doq^{(I} N_2^i \rd_i Q^{J)} } + 2 \alpha_1  \lrp{ \doq^{(I} \rd^i\beta_1 \rd_i Q^{J)} + \dop_0^{(I} N_2^i \rd_i Q^{J)}   } \\
        &\qquad\qquad  + \frac{1}{2}  \lrp{ 3 \alpha _1^2 - 2 \alpha_2} \lrp{ \doq^I \doq^J - 2\, \dop_0^{(I} \rd^i\beta_1 \rd_i Q^{J)}  } + \lrp{ -4 \alpha _1^3  +6 \alpha _1 \alpha _2} \dop_0^{(I} \doq^{J)} \\
        &\qquad\qquad  + \frac{1}{2}  \lrp{ 5 \alpha _1^4 -12 \alpha _1^2 \alpha _2+3 \alpha_2^2 }
         \dop_0^I \dop_0^J \,,\\
    X^{IJ}_3 &= -  \lrp{ \doq^{(I} \rd^i\beta_1 \rd_i Q^{J)} + \dop_0^{(I} N_2^i \rd_i Q^{J)} }  - \alpha_1 \lrp{ \doq^I \doq^J - 2\, \dop_0^{(I} \rd^i\beta_1 \rd_i Q^{J)} } \\
    &\qquad\qquad + \lrp{ 3 \alpha _1^2 - 2 \alpha_2} \dop_0^{(I} \doq^{J)}  + \lrp{ -2 \alpha _1^3  +3 \alpha _1 \alpha _2} \dop_0^I
    \dop_0^J \,,\\
        X^{IJ}_2 &=  -\frac{1}{2a^2} \rd_i Q^I \rd_i Q^J  + \frac{1}{2} \lrsb{ \lrp{\doq^I \doq^J - 2\, \dop_0^{(I} N^i_{1} \rd_i Q^{J)}} - 4\, \alpha_1 \, \dop_0^{(I} \doq^{J)} +
        \lrp{ 3 \alpha _1^2 - 2 \alpha_2}\, \dop_0^I \dop_0^J } \,.
 }
In this work, at leading-order
    \ea{
        X^{IJ}_1 &\simeq \dop_0^{(I} \doq^{J)} \,,\\
        X^{IJ}_2 &\simeq -\frac{1}{2a^2} \rd_i Q^I \rd_i Q^J +
        \frac{1}{2} \doq^I \doq^J \,.
    }

\section{Interaction Hamiltonian}{\label{appsec_Hamiltonian}}

\subsection{Derivation of interaction Hamiltonian}
In this appendix, we show the explicit steps to get the interaction
Hamiltonian. We start from the following leading-order Lagrangian
(2nd, 3rd and 4th-order respectively):
    \ea{
        \mathcal{L}_2 &=  \frac{1}{2} \mathcal{K}_{mn} \doq_m \doq_n - \frac{1}{2a^2} \delta_{mn} \rd_i Q_m \rd_i
        Q_n \,,\\
        \mathcal{L}_3 &=  \frac{1}{2} \Xi_{mnl} \doq_m \doq_n \doq_l  - \frac{1}{2a^2} \Upsilon_{mnl}\, \doq_m \rd_iQ_n \rd_iQ_l
        \,,\\
        \mathcal{L}_4 &=  \Gamma_{mnpq}\, \doq_m  \doq_n \doq_p \doq_q - \frac{1}{4a^2} \Theta_{mnpq}\, \doq_m \doq_n \rd_i Q_p \rd_i Q_q + \frac{1}{8a^4}\Omega_{mnpq}\, \rd_i Q_m \rd_i Q_n \rd_j Q_p \rd_j
        Q_q \,.
    }
The conjugate momentum of $Q_m$ is simply
    \ea{{\label{app_pi}}
        \pi_m &\equiv \pd{\mathcal{L}}{\doq_m} \\
        &= \mathcal{K}_{mn} \doq_n + \frac{3}{2} \Xi_{mnl} \doq_n
        \doq_l - \frac{1}{2a^2} \Upsilon_{mnl} \rd_iQ_n \rd_i Q_l +
        4\Gamma_{mnpq} \doq_n \doq_p \doq_q - \frac{1}{2a^2}
        \Theta_{mnpq} \doq_n \rd_iQ_p \rd_iQ_q \,.
    }
In order to get the Hamiltonian, we must solve $\doq_m$ in terms of
its conjugate momentum $\pi_m$. In general, it is a complicated
task. However, this can be done perturbatively. We make the ansatz:
    \eq{{\label{app_doq_pi_ansatz}}
        \doq_m \equiv \lambda^{(1)}_{mn}\, \pi_n +
        \lambda^{(2)}_{mnp}\, \pi_n \pi_p + \lambda^{(3)}_{mnpq}\,\pi_n \pi_p \pi_q
        + \mathcal{O}(\pi^4) \,,
    }
where $\lambda^{(i)}$'s are of order unity quantities, which we
should determine in the following. Note that in
(\ref{app_doq_pi_ansatz}), for our purpose, higher-order terms  are
not needed.

In order to determine $\lambda^{(i)}$'s in
(\ref{app_doq_pi_ansatz}), the strategy is that, plug
(\ref{app_doq_pi_ansatz}) into (\ref{app_pi}), and solve
$\lambda^{(i)}$'s order-by-order.
    \ea{{\label{app_pi_coeff}}
        \pi_m &\equiv \mathcal{K}_{mn} \lrp{ \lambda^{(1)}_{np}\, \pi_p + \lambda^{(2)}_{npq}\,\pi_p\pi_q + \lambda^{(3)}_{npql}\, \pi_p \pi_q \pi_l
        } + \frac{3}{2} \Xi_{mnl} \lrp{ \lambda^{(1)}_{np} \lambda^{(1)}_{lr}\, \pi_p \pi_r + 2 \lambda^{(1)}_{np} \lambda^{(2)}_{lrs}\, \pi_p \pi_r\pi_s
        } \\
        &\qquad -\frac{1}{2a^2} \Upsilon_{mnl}\, \rd_i Q_n \rd_i Q_l
        + 4 \Gamma_{mnpq}\, \lambda^{(1)}_{nl} \lambda^{(1)}_{pr}
        \lambda^{(1)}_{qs}\, \pi_l\pi_r \pi_s - \frac{1}{2a^2}
        \Theta_{mnpq}\, \lambda^{(1)}_{nl}\,\pi_{l} \rd_i Q_p
        \rd_iQ_q \,.
    }
At first-order, in order to satisfy (\ref{app_pi_coeff}), we have
    \eq{
        \pi_m \equiv \mathcal{K}_{mn} \, \lambda^{(1)}_{np}\, \pi_p\,,
    }
this gives
    \eq{
        \lambda^{(1)}_{mn} = \mathcal{K}^{-1}_{mn} \,.
    }
At the second-order, (\ref{app_pi_coeff}) gives
    \ea{
        \lambda^{(2)}_{spq}\, \pi_p\pi_q &\equiv
        -\frac{3}{2} \Xi_{mnl} \mathcal{K}^{-1}_{sm} \mathcal{K}^{-1}_{np}
        \mathcal{K}^{-1}_{lr}\, \pi_p\pi_r + \frac{1}{2a^2}
        \Upsilon_{mnl} \mathcal{K}^{-1}_{sm} \rd_iQ_n \rd_i Q_l  \,.
    }
Fortunately, we do not need to solve $\lambda^{(2)}_{npq}$
explicitly, the above relation is enough for our purpose. Similarly,
at the third-order we have
        \ea{
        &\quad\;  \lambda^{(3)}_{spql}\, \pi_p\pi_q
        \pi_l \\
        &\equiv  -3 \Xi_{mnl} \mathcal{K}^{-1}_{ms} \mathcal{K}^{-1}_{np}
        \lambda^{(2)}_{lrs}\, \pi_p \pi_r \pi_s - 4 \Gamma_{mnpq} \mathcal{K}^{-1}_{ms}
        \mathcal{K}^{-1}_{nl} \mathcal{K}^{-1}_{pr}
        \mathcal{K}^{-1}_{qs} \, \pi_l \pi_r \pi_s + \frac{1}{2a^2}
        \Theta_{mnpq} \mathcal{K}^{-1}_{ms} \mathcal{K}^{-1}_{nl} \, \pi_l \rd_i Q_p \rd_i
        Q_q  \,.
    }

The fourth-order part of the Hamiltonian density is
    \eq{{\label{app_H4}}
        \mathcal{H}_4 \equiv \pi_m \, \lambda^{(3)}_{mnpq}\, \pi_n
        \pi_p \pi_q - \mathcal{L}_{(4)} \,,
    }
where $\mathcal{L}_{(4)}$ is the fourth-order part of the Lagrangian
in terms of $Q_m$ and $\pi_m$ (note that $\mathcal{L}_2$,
$\mathcal{L}_3$ and $\mathcal{L}_4$ all contribute to
$\mathcal{L}_{(4)}$):
    \ea{
        \mathcal{L}_{(4)} &\equiv \mathcal{K}_{mn}\,
        \lambda^{(1)}_{np}\, \lambda^{(3)}_{mrst}\, \pi_p \pi_r
        \pi_s \pi_t + \frac{1}{2} \mathcal{K}_{mn}\,
        \lambda^{(2)}_{npq} \lambda^{(2)}_{mrs}\, \pi_p \pi_q \pi_r
        \pi_s + \frac{3}{2} \Xi_{mnl} \, \lambda^{(2)}_{mpq}\,
        \lambda^{(1)}_{nr} \lambda^{(1)}_{ls}\, \pi_p \pi_q \pi_r
        \pi_s \\
        &\qquad - \frac{1}{2a^2} \Upsilon_{mnl}\,
        \lambda^{(2)}_{mpq}\, \pi_p \pi_q \rd_i Q_n \rd_i Q_l +
        \Gamma_{mnpq}\, \lambda^{(1)}_{mr} \lambda^{(1)}_{ns} \lambda^{(1)}_{pu}
        \lambda^{(1)}_{qv}\, \pi_r\pi_s\pi_u\pi_v \\
        &\qquad - \frac{1}{4a^2}
        \Theta_{mnpq}\, \lambda^{(1)}_{mr} \lambda^{(1)}_{ns}\,
        \pi_r\pi_s \rd_i Q_p \rd_i Q_q + \frac{1}{8a^4}
        \Omega_{mnpq}\, \rd_i Q_m \rd_i Q_n \rd_j Q_p \rd_j Q_q \,.
    }
Thus,
    \ea{
        \mathcal{H}_4 &=  - \frac{1}{2} \mathcal{K}_{mn}\,
        \lambda^{(2)}_{npq} \lambda^{(2)}_{mrs}\, \pi_p \pi_q \pi_r
        \pi_s - \frac{3}{2} \Xi_{mnl} \, \lambda^{(2)}_{mpq}\,
        \mathcal{K}^{-1}_{nr} \mathcal{K}^{-1}_{ls}\, \pi_p \pi_q \pi_r
        \pi_s \\
        &\qquad + \frac{1}{2a^2} \Upsilon_{mnl}\,
        \lambda^{(2)}_{mpq}\, \pi_p \pi_q \rd_i Q_n \rd_i Q_l -
        \Gamma_{mnpq}\, \mathcal{K}^{-1}_{mr} \mathcal{K}^{-1}_{ns} \mathcal{K}^{-1}_{pu}
        \mathcal{K}^{-1}_{qv}\, \pi_r\pi_s\pi_u\pi_v \\
        &\qquad + \frac{1}{4a^2}
        \Theta_{mnpq}\, \mathcal{K}^{-1}_{mr} \mathcal{K}^{-1}_{ns}\,
        \pi_r\pi_s \rd_i Q_p \rd_i Q_q - \frac{1}{8a^4}
        \Omega_{mnpq}\, \rd_i Q_m \rd_i Q_n \rd_j Q_p \rd_j Q_q \,.
    }

\subsection{Interaction Hamiltonian in the interaction picture}

Now we go to the ``interaction picture", where all fields and
momenta are ``free". The free (quadratic) part Hamiltonian density
reads
    \ea{
        \mathcal{H}_0^{\I} &= \pi_m^{\I} \lambda^{(1)}_{mn} \pi_n^{\I} -
        \mathcal{L}_{(2)} = \frac{1}{2} \mathcal{K}^{-1}_{mn}\, \pi_m^{\I} \pi_n^{\I} + \frac{1}{2a^2} \delta_{mn} \rd_i Q_m^{\I} \rd_i Q_n^{\I} \,.
    }
Here we explicitly write superscript ``$\I$", which denotes
interaction picture quantities.
 In free theory, $\doq_m^{\I}$
is related with $\pi_m^{\I}$ by (remember that $\mathcal{H}_0^{\I}
\equiv \mathcal{H}_0$)
    \eq{
        \doq_m^{\I} \equiv \pd{\mathcal{H}_0}{\pi_m^{\I}} =
        \mathcal{K}^{-1}_{mn} \pi_n^{\I} \,,\qquad\qquad\textrm{i.e.}\qquad
        \pi_{m}^{\I} = \mathcal{K}_{mn} \doq_n^{\I} \,.
    }
Now our task is to plug $\pi_m^{\I}$ back into the fourth-order
Hamiltonian (\ref{app_H4}), to get its ``interaction picture form"
$\mathcal{H}_4^{\I}$. After a straightforward but rather tedious
calculation, we get
    \ea{{\label{app_H4_final}}
        \mathcal{H}_4^{\I} &= \doq_m^{\I} \doq_n^{\I} \doq_p^{\I} \doq_q^{\I} \lrsb{ \frac{9}{8} \mathcal{K}^{-1}_{rs}\Xi_{rmn} \, \Xi_{spq}  - \Gamma_{mnpq}
        } + \frac{1}{a^2}\doq_m^{\I} \doq_n^{\I} \rd_i Q_p^{\I} \rd_i Q_q^{\I} \lrp{ \frac{1}{4}
        \Theta_{mnpq}- \frac{3}{4} \mathcal{K}^{-1}_{rs}\Upsilon_{rpq}\, \Xi_{smn}
        } \\
        &\qquad + \frac{1}{a^4} \rd_i Q_m^{\I} \rd_i Q_n^{\I} \rd_j Q_p^{\I} \rd_j Q_q^{\I}  \lrp{ \frac{1}{8}
        \mathcal{K}^{-1}_{rs} \Upsilon_{rmn} \,\Upsilon_{spq}  -  \frac{1}{8}
        \Omega_{mnpq} } \,.
    }
We should keep in mind that the above $\mathcal{H}_4^{\I}$ is the
fourth-order interaction Hamiltonian in the ``interaction picture".
In this work, we work in operator formalism in the interaction
picture, thus, we use (\ref{app_H4_final}) as our staring point.

\section{Coefficients in $\mathcal{H}_4$}\label{appsec_H4_coeff}

The various coefficients in the 4th-order interaction Hamiltonian
density (\ref{H4_sigma}), (\ref{H4_s}) and (\ref{H4_cross}) are
    \ea{
        \Gamma_{\sigma} &\equiv \frac{ -H^2 \epsilon  (3 \lambda +10 \Pi )+54 \lambda ^2 c_a^2 }{24 H^6 \epsilon ^3} \,,\\
        \Gamma_s &\equiv\frac{ c_a^4 \left(7-4 c_e^2\right){}^2+4 \left(c_e^2+c_e^4\right)+c_a^2 \left(-13-20 c_e^2+16 c_e^4\right) }{16 H^2 \epsilon  c_a^2 c_e^4} \,,\\
        \Gamma_c &\equiv \frac{1}{4 H^4 \epsilon ^2 c_a^4 c_e^2}
        \left[ H^2 \epsilon  \left(-1+c_e^2\right){}^2+3 \lambda  c_a^6 \left(-7+4 c_e^2\right)+H^2 \epsilon  c_a^2 \left(5-8 c_e^2+4 c_e^4\right) \right.\\
        &\qquad\qquad\qquad\qquad \left. +c_a^4 \left(3 H^2 \epsilon +\left(-8 H^2 \epsilon +6 \lambda \right) c_e^2+4 H^2 \epsilon  c_e^4\right)
        \right] \,,}
    \ea{
        \Theta_{\sigma} &\equiv  \frac{ -H^2 \epsilon +H^2 \epsilon  c_a^2+3 \lambda  c_a^4  }{4 H^4 \epsilon ^2 c_a^2} \,,\\
        \Theta_{\sigma s} &\equiv  \frac{ H^2 \epsilon  c_a^2+H^2 \epsilon  \left(-2+c_e^2\right)-3 \lambda  c_a^4 \left(-2+c_e^2\right)  }{4 H^4 \epsilon ^2 c_a^2 c_e^2} \,,\\
        \Theta_{s\sigma} &\equiv  \frac{ 5 c_a^2-2 c_e^2+c_a^4 \left(-7+4 c_e^2\right) }{8 H^2 \epsilon  c_a^2 c_e^2} \,,\\
        \Theta_{s} &\equiv  \frac{ -2 c_e^4+c_a^2 \left(2+9 c_e^2-6 c_e^4\right)+c_a^4 \left(-14+15 c_e^2-4 c_e^4\right) }{8 H^2 \epsilon  c_a^2 c_e^4} \,,\\
        \Theta_{c} &\equiv - \frac{\left(-1+c_e^2\right) \left(c_e^2+2 c_a^2 \left(-1+c_e^2\right)\right) }{2 H^2 \epsilon  c_a^2 c_e^2}
        \,,}
    \ea{
        \Omega_{\sigma} &\equiv  \frac{ -1+c_a^2}{16 H^2 \epsilon } \,,\qquad\qquad \Omega_{s} \equiv \frac{ c_a^2 \left(-2+c_e^2\right)^2+c_e^2 \left(-4+3 c_e^2\right) }{16 H^2 \epsilon  c_e^4} \,,\\
        \Omega_{\sigma s} &\equiv  -\frac{ c_e^2+c_a^2 \left(-2+c_e^2\right) }{8 H^2 \epsilon  c_e^2} \,,\qquad\qquad \Omega_{c} \equiv  \frac{ -1+c_e^2  }{4 H^2 \epsilon }  \,.
    }

\section{Explicit Expressions for $\lrab{\qsg\qsg\qs\qs}_i$}\label{appsec_2a2s}

\ea{
    &\quad\; \lrab{ \qsg \qsg \qs\qs}_2 \\
    &\equiv (2\pi)^3 \delta^3(\vk_{1234}) \lrp{ -2\Re} \lrsb{ 4 \Theta_{\sigma s}\, i \int_{-\infty}^{\eta_{\ast}} d\eta\, \lrp{-\vk_3 \cdot \vk_4} \od{}{\eta} G_{k_1}(\eta_{\ast},\eta) \od{}{\eta} G_{k_2}(\eta_{\ast},\eta)  F_{k_3}(\eta_{\ast},\eta)  F_{k_4}(\eta_{\ast},\eta)
    } \\
    &=  (2\pi)^3 \delta^3(\vk_{1234}) \Theta_{\sigma s}  \lrp{-\vk_3 \cdot
    \vk_4} \frac{H^8 c_a^2 \, k_1^2 k_2^2 }{ c_e^2\, \tilde{K}^3  \prod_i^4 k_i^3 }  \lrsb{  1 +  \frac{12\, c_e^2 k_3 k_4 }{ \tilde{K}^2 }+ \frac{ 3\,  c_e \left(k_3+k_4\right) }{ \tilde{K}  }
    } \\
}

\ea{
    &\quad\; \lrab{ \qsg \qsg \qs\qs}_3 \\
    &\equiv (2\pi)^3 \delta^3(\vk_{1234}) \lrp{ -2\Re} \lrsb{ 4 \Theta_{s \sigma}\, i \int_{-\infty}^{\eta_{\ast}} d\eta\, \lrp{-\vk_1 \cdot \vk_2}  G_{k_1}(\eta_{\ast},\eta) G_{k_2}(\eta_{\ast},\eta) \od{}{\eta} F_{k_3}(\eta_{\ast},\eta) \od{}{\eta} F_{k_4}(\eta_{\ast},\eta)
    } \\
    &= (2\pi)^3 \delta^3(\vk_{1234}) \Theta_{s \sigma}  \lrp{-\vk_1 \cdot
    \vk_2} \frac{H^8 c_e^2 \, k_3^2 k_4^2 }{ c_a^2\, \tilde{K}^3  \prod_i^4 k_i^3 }  \lrsb{  1 +  \frac{12\, c_a^2 k_1 k_2 }{ \tilde{K}^2 }+ \frac{ 3 \, c_a \left(k_1+k_2\right) }{ \tilde{K}  }
    } \\
}

\ea{
    &\quad\; \lrab{ \qsg \qsg \qs\qs}_4 \\
    &\equiv (2\pi)^3 \delta^3(\vk_{1234}) \lrp{ -2\Re} \lrsb{ \Theta_{c}\, i \int_{-\infty}^{\eta_{\ast}} d\eta\, \lrp{-\vk_2 \cdot \vk_4}  \od{}{\eta}G_{k_1}(\eta_{\ast},\eta) G_{k_2}(\eta_{\ast},\eta) \od{}{\eta} F_{k_3}(\eta_{\ast},\eta)  F_{k_4}(\eta_{\ast},\eta)
    +\textrm{3 perms}} \\
    &= (2\pi)^3 \delta^3(\vk_{1234}) \Theta_c \frac{ H^8 }{ 4\,\tilde{K}^3 \prod_i^4 k_i^3} \lrsb{ \lrp{-\vk_2 \cdot \vk_4} k_1^2 k_3^2 \lrp{ 1 + \frac{12 c_a c_e k_2 k_4}{\tilde{K}^2} + \frac{ 3 \left(c_a k_2+c_e k_4\right)   }{\tilde{K} } } +\textrm{3 perms} }\\
} The other permutation is $(\vk_2,\vk_3)$, $(\vk_1,\vk_3)$ and
$(\vk_1,\vk_4)$ combination.

 \ea{
    &\quad\; \lrab{ \qsg \qsg \qs\qs}_5 \\
    &\equiv (2\pi)^3 \delta^3(\vk_{1234})
    \lrp{-2\Re} \lrsb{ 4 \Omega_{\sigma s} \lrp{ \vk_1 \cdot \vk_2 } \lrp{ \vk_3 \cdot \vk_4 } \, i\int_{-\infty}^{\eta_{\ast}}d\eta\, G_{k_1}(\eta_{\ast},\eta) G_{k_2}(\eta_{\ast},\eta) F_{k_3}(\eta_{\ast},\eta)
    F_{k_4}(\eta_{\ast},\eta)} \\
    &=  (2\pi)^3 \delta^3(\vk_{1234}) \Omega_{\sigma s} \frac{ \lrp{ \vk_1 \cdot \vk_2 } \lrp{ \vk_3 \cdot \vk_4
    } }{\tilde{K}} \lrp{ -\frac{H^8}{ c_a^2 c_e^2  \prod_i^4 k_i^3  }  } \\
    &\qquad\times \left[  1 + \frac{ 12 c_a^2 c_e^2 \prod_i^4 k_i }{\tilde{K}^4}  + \frac{3}{ \tilde{K}^3} c_a c_e \left(c_e \left(k_1+k_2\right) k_3 k_4+c_a k_1 k_2 \left(k_3+k_4\right)\right) \right. \\
    &\qquad\qquad\qquad\qquad \left.  + \frac{1}{\tilde{K}^2 }  \left(c_a^2 k_1 k_2+c_e^2 k_3 k_4+c_a c_e \left(k_1+k_2\right) \left(k_3+k_4\right)\right)
    \right]
}

\ea{
    &\quad\; \lrab{ \qsg \qsg \qs\qs}_6 \\
    &\equiv (2\pi)^3 \delta^3(\vk_{1234})
    \lrp{-2\Re} \lrsb{ 2 \Omega_{c} \lrp{ \vk_1 \cdot \vk_3 } \lrp{ \vk_2 \cdot \vk_4 } \, i\int_{-\infty}^{\eta_{\ast}}d\eta\, G_{k_1}(\eta_{\ast},\eta) G_{k_2}(\eta_{\ast},\eta) F_{k_3}(\eta_{\ast},\eta)
    F_{k_4}(\eta_{\ast},\eta) + \textrm{1 perm} } \\
    &= (2\pi)^3 \delta^3(\vk_{1234}) \Omega_{c}  \, \frac{ \lrp{ \vk_1 \cdot \vk_3 } \lrp{ \vk_2 \cdot \vk_4 }
    }{2\tilde{K}} \lrp{ -\frac{H^8}{ c_a^2 c_e^2  \prod_i^4 k_i^3  }  } \\
    &\qquad\times \left[ \left(  1 + \frac{ 12 c_a^2 c_e^2 \prod_i^4 k_i }{\tilde{K}^4}  + \frac{3}{ \tilde{K}^3} c_a c_e \left(c_e \left(k_1+k_2\right) k_3 k_4+c_a k_1 k_2 \left(k_3+k_4\right)\right) \right.\right. \\
    &\qquad\qquad \left. \left.  + \frac{1}{\tilde{K}^2 }  \left(c_a^2 k_1 k_2+c_e^2 k_3 k_4+c_a c_e \left(k_1+k_2\right) \left(k_3+k_4\right)\right)
    \right) + \textrm{1 perm} \right] \,.
}



\end{document}